\documentclass{ieeeaccess}
\usepackage{cite}
\usepackage{amsmath,amssymb,amsfonts}
\usepackage{algorithmic}
\usepackage{graphicx}
\usepackage{textcomp}
\def\BibTeX{{\rm B\kern-.05em{\sc i\kern-.025em b}\kern-.08em
    T\kern-.1667em\lower.7ex\hbox{E}\kern-.125emX}}
\usepackage{url}
\usepackage{hyperref}
\begin{document}
\history{PREPRINT, submitted to IEEE Access}
\doi{PREPRINT, submitted to IEEE Access}

\title{6-Layer Model for a Structured Description and Categorization of Urban Traffic and Environment}
\author{\uppercase{M. Scholtes}\authorrefmark{1}, 
\uppercase{L. Westhofen\authorrefmark{2}, L. R. Turner \authorrefmark{3}, K. Lotto\authorrefmark{3}, M. Schuldes\authorrefmark{1}, H. Weber\authorrefmark{1}, N. Wagener\authorrefmark{1}, C. Neurohr\authorrefmark{2}, M. H. Bollmann\authorrefmark{3}, F. Körtke\authorrefmark{3}, J. Hiller\authorrefmark{1}, M. Hoss\authorrefmark{1}, J. Bock\authorrefmark{4} and L. Eckstein}\authorrefmark{1}
}
\address[1]{Institute for Automotive Engineering (ika), RWTH Aachen University, 52074 Aachen, Germany}
\address[2]{OFFIS e.V., 26121 Oldenburg, Germany}
\address[3]{ZF Friedrichshafen AG, 88046 Friedrichshafen, Germany}
\address[4]{fka GmbH, 52074 Aachen, Germany}
\tfootnote{The research leading to these results is funded by the German Federal Ministry for Economic Affairs and Energy within the project `VVM - Verification \& Validation Methods for Automated Vehicles Level 4 and 5'.}

\markboth
{Scholtes \headeretal: 6-Layer Model for a Structured Description and Categorization of Urban Traffic and Environment}
{Scholtes \headeretal: 6-Layer Model for a Structured Description and Categorization of Urban Traffic and Environment}

\corresp{Corresponding author: M. Scholtes (e-mail: maike.scholtes@ika.rwth-aachen.de).}

\begin{abstract}
Verification and validation of automated driving functions impose large challenges. Currently, scenario-based approaches are investigated in research and industry, aiming at a reduction of testing efforts by specifying safety relevant scenarios. To define those scenarios and operate in a complex real-world design domain, a structured description of the environment is needed. Within the PEGASUS research project, the 6-Layer Model (6LM) was introduced for the description of highway scenarios. This paper refines the 6LM and extends it to urban traffic and environment. As defined in PEGASUS, the 6LM provides the possibility to categorize the environment and, therefore, functions as a structured basis for subsequent scenario description. The model enables a structured description and categorization of the general environment, without incorporating any knowledge or anticipating any functions of actors. Beyond that, there is a variety of other applications of the 6LM, which are elaborated in this paper. The 6LM includes a description of the road network and traffic guidance objects, roadside structures, temporary modifications of the former, dynamic objects, environmental conditions and digital information. The work at hand specifies each layer by categorizing its items. Guidelines are formulated and explanatory examples are given to standardize the application of the model for an objective environment description. In contrast to previous publications, the model and its design are described in far more detail. Finally, the holistic description of the 6LM presented includes remarks on possible future work when expanding the concept to machine perception aspects.
\end{abstract}


\titlepgskip=-15pt

\maketitle

\section{Introduction}
\label{sec:Introduction}
\PARstart{A}{s} automated driving (AD) constantly increases in importance \cite{b1}, a large challenge faced when implementing (highly) automated driving (HAD) functions is testing and validation of such functions. In \cite{b2}, the issue arising when trying to validate HAD through real-world drives, the so-called `approval trap', is described. The resulting amount of kilometers that needs to be driven for this distance-based statistical validation approach is not feasible due to time and cost reasons. This motivates the idea of scenario-based verification and validation, in which specific safety-relevant scenarios guide the testing process \cite{b3}.

In order to use the scenario-based method for automated vehicles operating in an open context \cite{b4}, i.e. an unstructured real-world operational design domain, a sufficiently complete description of the environment is needed. To decrease the complexity and provide a structured method the environment can be described by utilizing the 6-Layer Model (6LM) which was already introduced in previous work (see Section \ref{sec:RelatedWork}). Within the German research project PEGASUS and the context of defining highway scenarios \cite{b5,b6}, the concept was applied to separate relevant aspects of the environment description into different layers that are built upon each other.

When extending the approach to urban use cases many other aspects need to be discussed. Although urban structures were already categorized before, see, in particular, \cite{b7}, they were – due to the different focus on highways – not further considered in PEGASUS and the following publications. Additionally, we could observe a rather discontinuous, even conflicting, evolution of the Layer Model. Entities and properties were shifted between layers and their naming was often changed such that numerous ambiguities exist today. For this reason, we see the need for a proper definition of the 6LM which will be given in this work to make the model a more accessible tool for structured environment description to researchers and safety engineers. The six layers presented in this paper heavily build upon existing literature as indicated in the definitions and examples of Section \ref{sec:Definition} and \ref{sec:Guidelines}. The classification of objects from previous work is used where appropriate, but clarified or adapted where needed with the goal to provide a single and consistent reference of the 6LM for all use cases.

The remainder of this work is structured as follows. An overview of previous publications dealing with environment descriptions based on a layered model is given in Section \ref{sec:RelatedWork}. Furthermore, in Section \ref{sec:Scope}, the scope of the refined model as well as some motivation on where it can be used is presented. In the following section, we extend the model by a detailed description of each layer including the naming and a comparison to previous definitions. We propose a categorization of relevant traffic entities and their properties into a high-level classification supported by interesting examples. The description of the model is followed by eight guidelines. Explanatory examples for each guideline reveal how to categorize different entities with their properties and provide justification for the definition of the layers. Subsequently, the presented definitions and guidelines are applied to a real-world example. The work is concluded by an outlook onto future work and a summary of the described approach.

\section{Related Work}
\label{sec:RelatedWork}
In order to understand the 6LM and its applications, it is important to look more closely on the definition of the term scenario. Definitions of scene and scenario are given in \cite{b8,b9} and in a DIN SAE Spec \cite{b10}. While a scene is described as a snapshot of the environment, a scenario describes the temporal development of those snapshots. Therefore, a scenario features a certain time span. This is important to note for the following description of the 6LM.

\begin{figure*}[th!]
\centering
\centerline{\includegraphics[width=0.95\textwidth, scale=0.75]{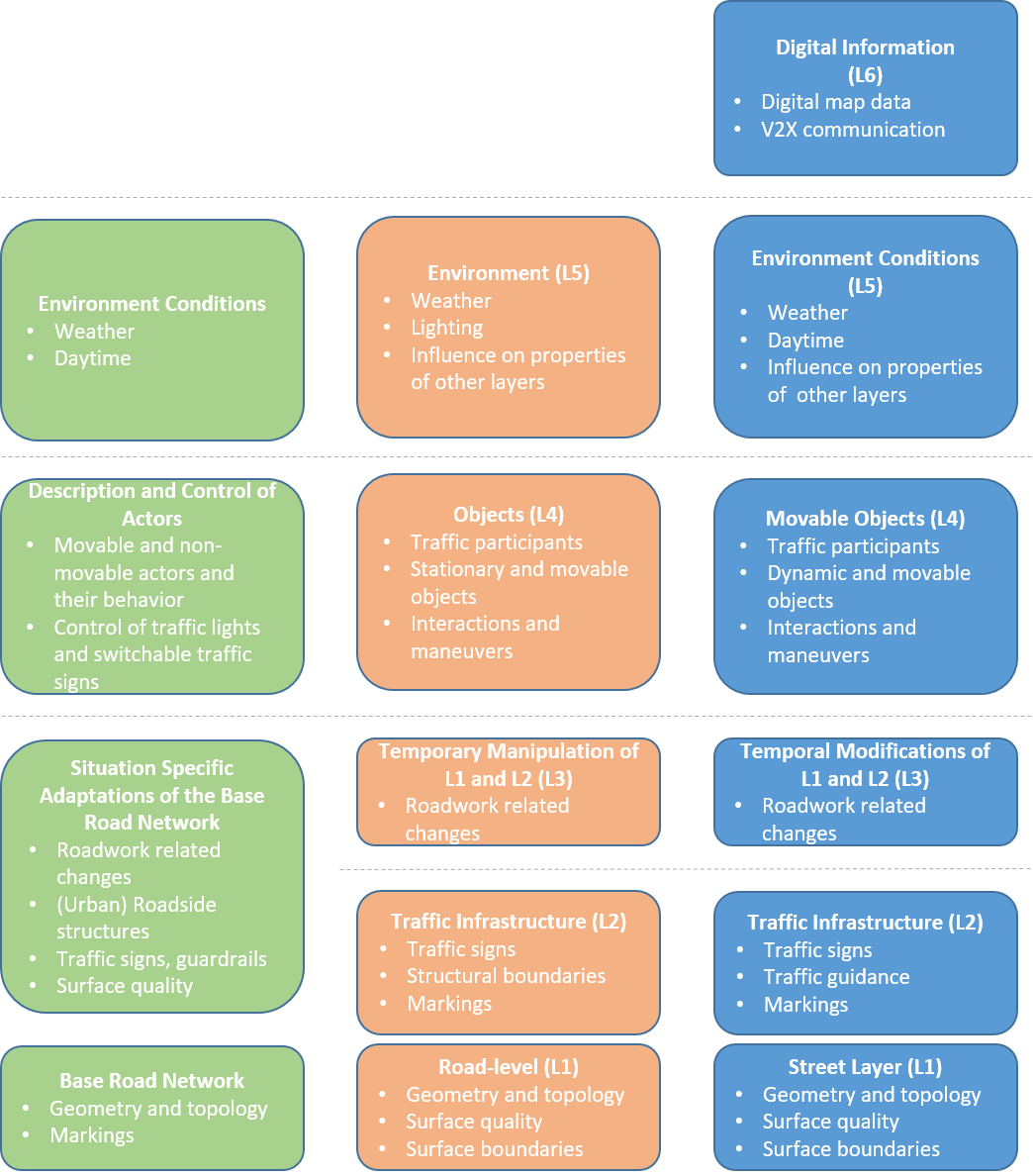}}
\caption{Historic development of the 6LM. Four layers of Schuldt \cite{b11} (left) compared to the five layers of Bagschik et al. \cite{b12} (center) and the six layers of Bock et al. \cite{b13} (right).}
\label{fig:history}
\end{figure*}

The concept of using a layered model to structure an environment description for scenes and scenarios was first introduced in \cite{b7}. In his Ph.D. thesis \cite{b11}, the lead author of the previous publication refined and adapted the model. There, the model featured four layers: The first layer for the base road network, the second layer for situation-specific adaptations of the road network, a third layer to describe the actors and their control and a last layer for environment conditions. Within this concept, only hierarchical higher ranked layers could influence lower ranked layers. All layers introduced in \cite{b11} can be found in Fig.~\ref{fig:history}, which provides an overview of the development of the 6LM. In \cite{b11} markings are included in the first layer while signs, guardrails and (urban) roadside structures are located in the second layer. In this reference, the second layer performs situation-specific adaptations to the road network required for special applications and automated driving functions. This includes the placement of different signs and safety structures as well as buildings and street lamps to construct various environments. Roadwork related changes are also mentioned in this context. Therefore, Layer~2 of \cite{b11} is a combination of Layer~2 and Layer~3 of subsequent work that performs a more distinctive classification.

Within PEGASUS \cite{b6} the concept of \cite{b11} was taken up for the use case of highway scenarios. Subsequently, a fifth \cite{b12} and a sixth layer \cite{b13} were introduced, resulting in the 6LM discussed in this work. In \cite{b12} the former base road network is split into two different layers, namely the road-level and the traffic infrastructure, as to separate the road description and the traffic rules. The road-level in \cite{b12} only contains the layout of the road (geometry) and its topology while the layer for traffic infrastructure contains structural boundaries, traffic signs and markings (the latter in contrast to \cite{b11}). In a German publication \cite{b14}, which was published alongside the English version \cite{b12}, some small, but meaningful differences are present. E.g., in \cite{b14}, Layer~2 is named road equipment while in the English translation this is changed to traffic infrastructure. 

In all subsequent work (\cite{b12, b13, b14, b15}) Layer~3 is separated from the previous Layer~2 and describes solely temporary modifications of Layer~1 and Layer~2. This can, e.g., include modifications made when a construction site is present. In \cite{b12} Layer~4 is named `Objects'. It contains all static, dynamic and movable objects that are not already part of the traffic infrastructure. Furthermore, maneuvers and interactions are situated in Layer~4. In \cite{b14} this layer is named slightly different (`Movable Objects'), but contains the same objects. However, in contrast to \cite{b12, b14} does not mention that `Movable Objects' includes all potentially movable objects (static / stationary objects in \cite{b12}), i.e., also traffic participants that do currently not move. Layer~5, sticks with the previous definition of Layer~4 from \cite{b11} describing environmental conditions, such as weather.

The concept in \cite{b13} is consistent with the basic concept of \cite{b12} using the five layers (and renaming them slightly): Street layer, traffic infrastructure, temporal modifications of Layer~1 and Layer~2, movable objects and environment conditions. However, \cite{b13} introduces a sixth layer for digital information. This sixth layer is later renamed `Data and Communication' in \cite{b15}, featuring the same definition as in the previous work. As \cite{b13} focuses on the highway use case, structural objects along the road, such as buildings, are not mentioned explicitly. Furthermore, the naming of Layer~4 as `Movable Objects' makes a more definite classification. This latest version of the model will be used as a starting point for the work at hand.

How the 6LM, as described in \cite{b13}, can be used to develop a framework for scenario definition is shown in \cite{b15}. The work performed in \cite{b15} utilizes Layer~4 of the 6LM to define logical scenarios \cite{b10}, \cite{b16} for controlled-access highways.

\section{Scope and Motivation}
\label{sec:Scope}

\begin{figure*}[th]
\centering
\centerline{\includegraphics[width=0.8\textwidth, scale=0.75]{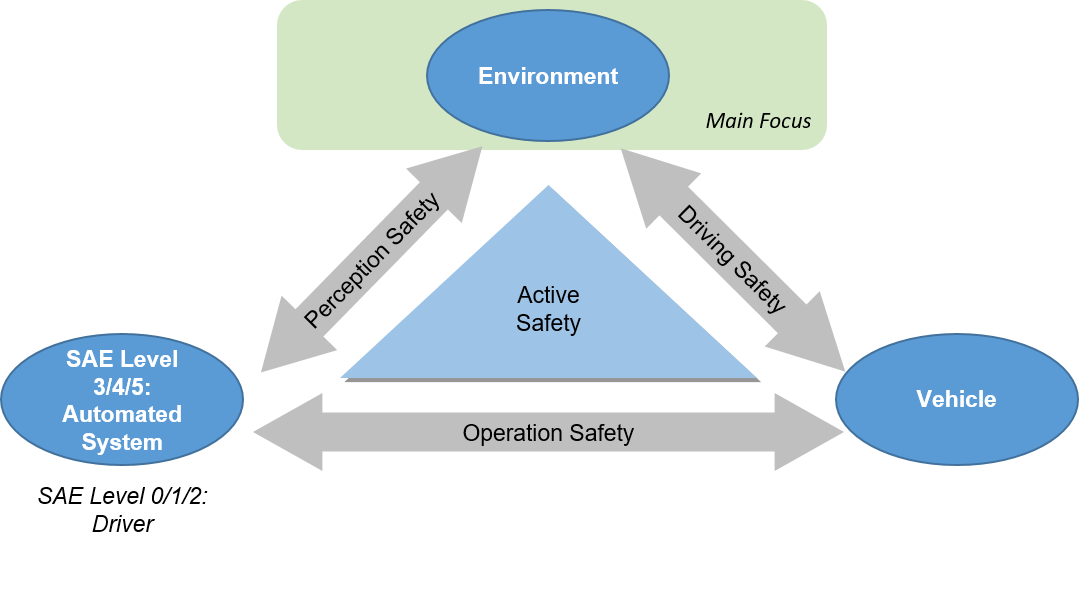}}
\caption{Schematic representation of the control loop between environment, automated system (for SAE Level 3/4/5 \cite{b19}) and vehicle and the focus of the 6LM.}
\label{fig:ControlLoop}
\end{figure*}

Given the control loop of environment, driver and vehicle as depicted in Fig.~\ref{fig:ControlLoop}, the main focus of the 6LM as proposed in previous work is on the structured categorization of the environment which naturally interacts with both, the driver and the vehicle. In the field of verification and validation, however, a series of additional applications of the 6LM exist. This extended scope will be motivated in this section.

In the following, the notions `entity', and `property' are heavily used. For the scope of this work, we refer to an entity as anything that exists, has existed or will exist \cite{b17}. The term `object' denotes a material entity only. This means that traffic participants, their maneuvers, and their intentions are all entities, but only the traffic participants themselves are objects. A property is an attribute of an entity that can be assigned some value, e.g., the position of a traffic sign, the size of a building, the visibility of a road marking, the velocity of a vehicle, the intensity of a precipitation, or the state of a traffic light. Besides properties, entities can also have relations to each other, i.e., linking them together. For instance, a car can be connected to another car by the relation `drives behind' \cite{b18}. Relations and classes of entities, e.g., the class of all vehicles, become more important in the context of ontologies.

Consider an engineer designing scenarios for the verification and validation of AD functions. This engineer can use the 6LM with its clear characterization of the environment, the entities and the properties as a basis for a scenario description. This holds independently of the utilized scenario description language itself, which could be natural, formal, or machine-readable. However, in order to allow for an automatic conversion between the scenario description and the language formats, we have to ensure that the structuring of the 6LM is compatible with the existing formats OpenDRIVE \cite{b20} and OpenSCENARIO \cite{b21}. Having this application in mind, the terms `environment' / `environment description' and `scenario' / `scenario description', respectively, will often be used interchangeably within this paper. This in turn means that the 6LM will mainly concentrate on the description of short time periods, i.e., the duration of a typical scenario. Note that clear and consistent rules for scenario design are particularly essential when reproducibility of scenarios is required, e.g. for testing campaigns. Focusing on hazardous scenarios, the 6LM was already applied to structure an environmental model used for iterative identification of those scenarios \cite{b22}.

Similarly, the 6LM can serve as a basis for a traffic domain ontology. In this paper, the authors examine how relevant domain entities and their properties can be categorized into a high-level classification of six layers, i.e., a flat, informally specified taxonomy. The canonical issue of detailing the single layers is part of ongoing research. Such an implementation of the 6LM can be done in a formal and digital ontology, e.g., by using the Web Ontology Language \cite{b23}, where the layers enable to classify entities by virtue of the well-defined categorization. 

Test engineers are already utilizing the model as a structured format for recording and analysis of measurement data to identify influencing factors on different layers and to finally derive a scenario concept on that basis. For this purpose, it is necessary to develop a holistic, i.e., comprehensive, and well-structured environment description with all relevant environment aspects assigned to the corresponding layers. Given that, the 6LM should also be suitable to describe environments in different settings -- urban, rural, and highway -- and on different abstraction levels - macro-, micro- and nanoscopic \cite{b24}.

The environment description should be unbiased and actor-independent at any time. Therefore, the 6LM must not anticipate any function of an actor or any properties of later steps, as they might, for instance, occur in scenario extraction. Regarding the control loop of environment, driver and vehicle (see Fig.~\ref{fig:ControlLoop}), this means that the 6LM is supposed to formulate a system-independent general environment description with an objective view on the traffic participants, without providing information on their expected behavior. In the same way, the 6LM should not contain any goals, values, or norms since the model is an `as it is'-description of the physically observable only. As such the 6LM is not suitable to describe situations as defined in \cite{b9}. The above holds independently from the introduction of automation where the driver of the control loop is replaced with the automated system, as schematically pictured in Fig.~\ref{fig:ControlLoop}. An outlook to possible new aspects of the 6LM introduced through automation such as machine perception can be found in Section \ref{sec:FutureWork}.

The need for a method to construct (virtual) environments was frequently experienced in the past. In order to close this gap and to support the generation of scenarios, the refinement of the 6LM focuses on a simple and unambiguous design. This also facilitates the comparison of scenarios and a subsequent reduction of given scenario sets. This identification of similarities and differences can be performed on the entire description, but in most cases, it might be advisable to perform it on single layers. The same holds for testing and debugging of system failures. Consider a behavior or motion planning function that can be tested with road network, traffic guidance objects, and dynamic objects, but in contrast to a perception function, without any interference through roadside structures and environmental conditions. A similar requirement on the layers' independence could be imposed by a simulation engineer who wants to execute simplified environment simulations on selected layers or full environment simulations on all layers, depending on the power of the simulation tool at hand. Furthermore, there are cases where only the content of single layers is intended for exchange between different stakeholders like OEMs and Tier-1s. There might be layers with common and such with more customized content. These layers should be preferably separated from each other and the layers should be as self-contained as possible. 

To conclude this section, we look at the 6LM from a different – highly intuitive – perspective: Imagine the 6LM as a kind of board game where Layer~1 up to Layer~3 serve as a base description of the board. Then, Layer~4 describes the actors with their behavior as well as comparable dynamic incidents on the board. Layer~5 and Layer~6 are somewhat separated from this description and can be imagined above the board, but with influences on the board and its actors. Finally, this metaphor also gives an indication for the ordering of the layers. The numbering of the layers is not to rank them by importance, but purely for structuring and categorization where one layer builds upon another.

\section{Definition of the Layers}
\label{sec:Definition}

This section introduces the different layers of the 6LM. For each layer, it provides an explanatory name and categories of objects belonging to that layer. 
Table \ref{tab:LayerOverview} provides an exemplary, incomplete, overview of the entities within the different layers.

\begin{table}[]
\caption{Layers of the 6LM including exemplary entities on the different layers}
\label{tab:LayerOverview}
\begin{tabular}{|p{90pt}|p{120pt}|}
\hline
\multicolumn{2}{|l|}{\textbf{Layer 6 - Digital Information}}                                                                                                                                                                                                    \\ \hline
\raisebox{-16mm}{\includegraphics[width=0.15\textwidth, height=25mm]{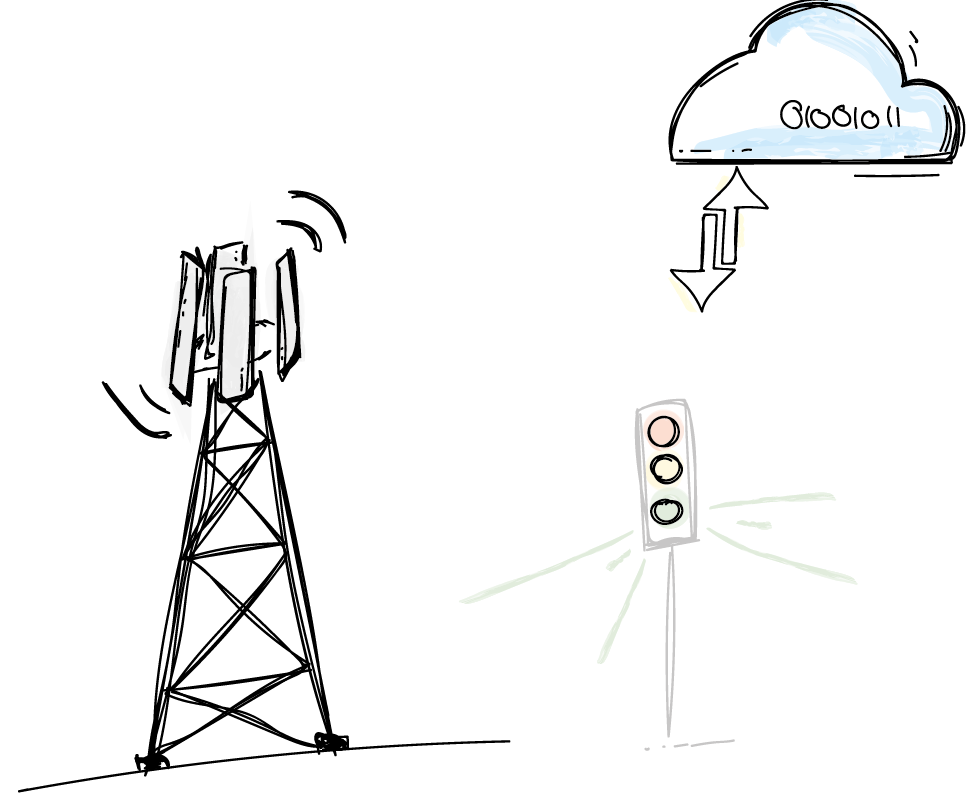}} & \begin{tabular}[c]{@{}l@{}}State of traffic lights and switchable\\ traffic signs\\ \\ V2X messages\\ \\ Cellular network coverage\end{tabular}                                                                                                                \\ \hline
\multicolumn{2}{|l|}{\textbf{Layer 5 - Environmental Conditions}}                                                                                                                                                                                               \\ \hline
\raisebox{-15mm}{\includegraphics[width=0.15\textwidth, height=25mm]{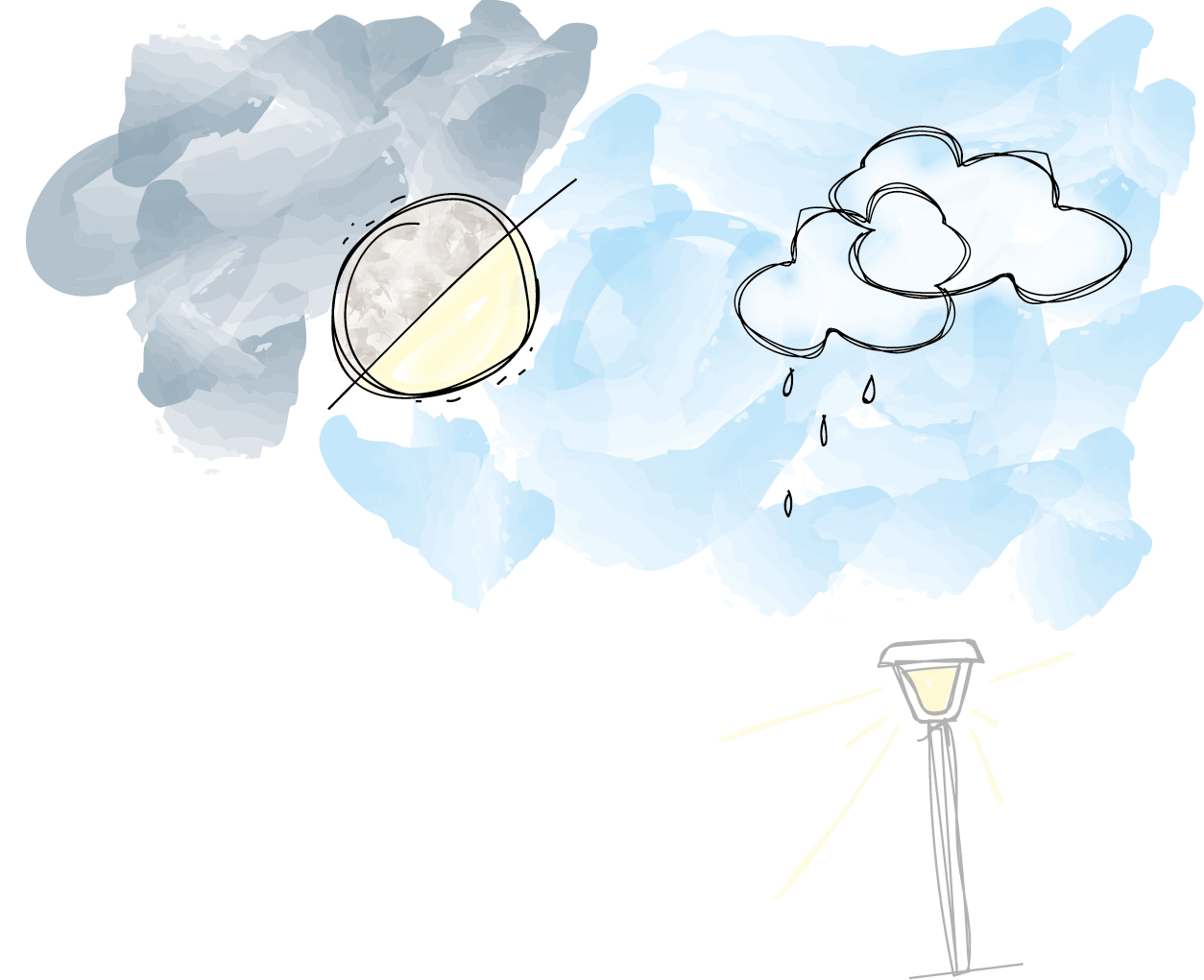}} & \begin{tabular}[c]{@{}l@{}}(Artificial) Illumination\\ \\ Precipitation\\ \\ Road weather (dry, wet, icy etc.)\\ \\ Wind\end{tabular}                                                                                                                        \\ \hline
\multicolumn{2}{|l|}{\textbf{Layer 4 - Dynamic Objects}}                                                                                                                                                                                                        \\ \hline
\raisebox{-8mm}{\includegraphics[width=0.18\textwidth, height=18mm]{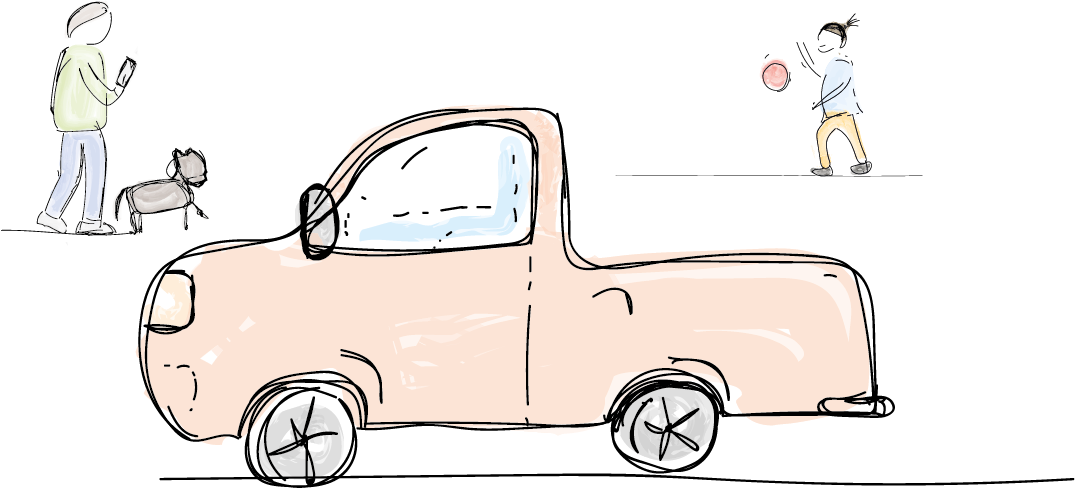}} & \begin{tabular}[c]{@{}l@{}}Vehicles (moving and non-moving)\\ \\ Pedestrians (moving and non-moving)\\ \\ Trailers\\ \\ Animals\\ \\ Trees falling over\\ (at the current point in time)\\ \\ Miscellaneous objects such as balls,\\ coke cans etc.\end{tabular} \\ \hline
\multicolumn{2}{|l|}{\textbf{Layer 3 - Temporary Modifications of L1 and L2}}                                                                                                                                                                                   \\ \hline
\raisebox{-15mm}{\includegraphics[width=0.15\textwidth, height=25mm]{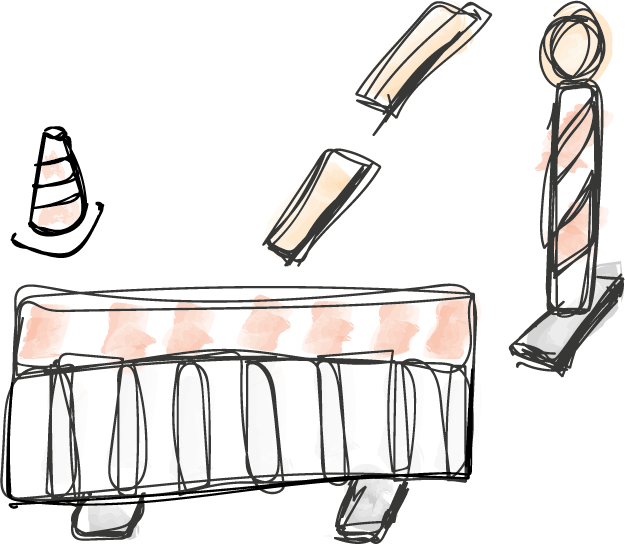}} & \begin{tabular}[c]{@{}l@{}}Roadwork signs\\ \\ Temporary markings\\ \\ Covered markings\\ \\ Fallen trees laying on the street\end{tabular}                                                                                                                  \\ \hline
\multicolumn{2}{|l|}{\textbf{Layer 2 - Roadside Structures}}                                                                                                                                                                                                    \\ \hline
\raisebox{-12mm}{\includegraphics[width=0.15\textwidth, height=25mm]{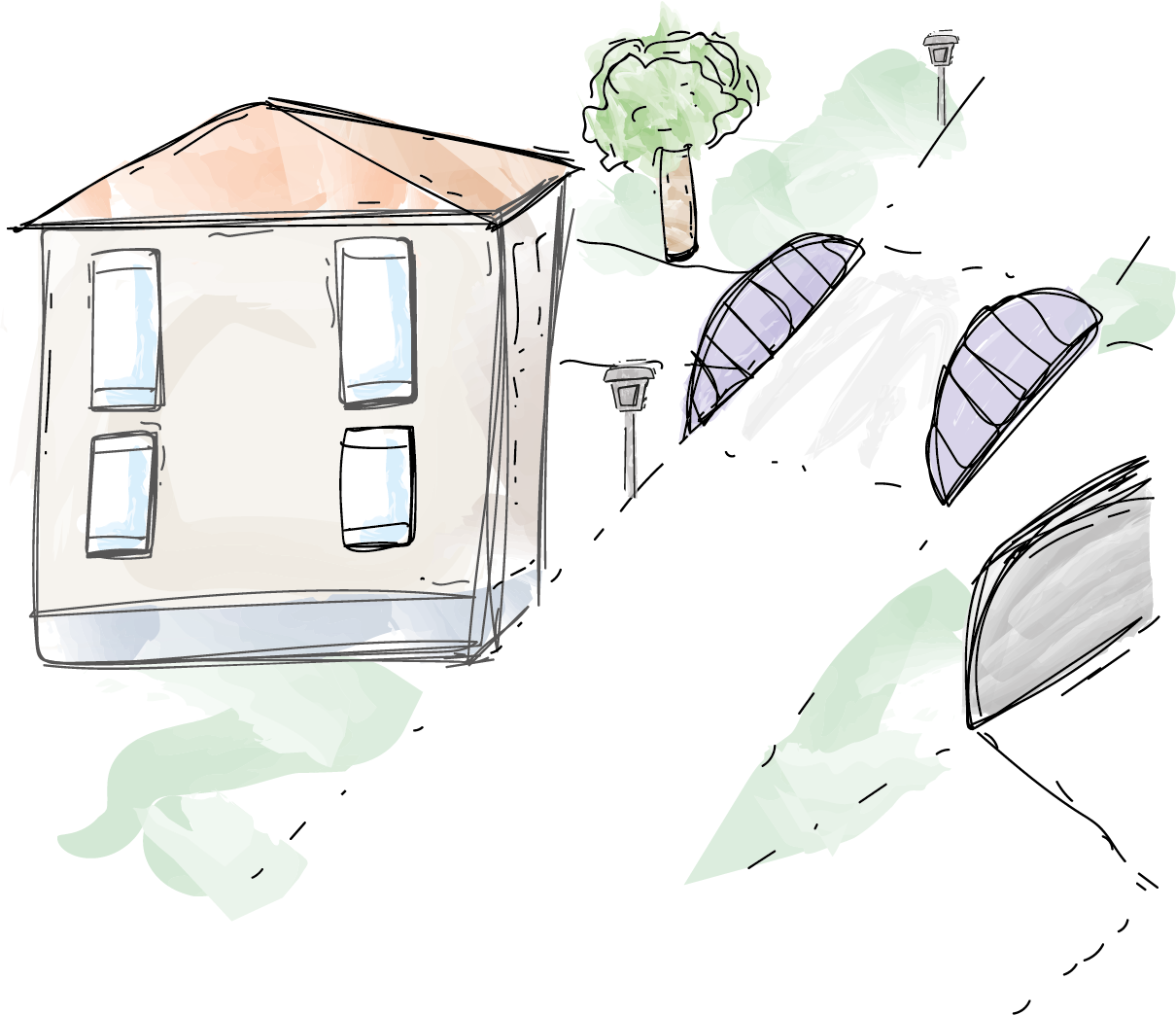}} & \begin{tabular}[c]{@{}l@{}}Buildings\\ \\ Vegetation\\ \\ Guardrails\\ \\ Street lamps\\ \\ Advertising boards and pillars\end{tabular}                                                                                                                      \\ \hline
\multicolumn{2}{|l|}{\textbf{Layer 1 - Road Network and Traffic Guidance Objects}}                                                                                                                                                                              \\ \hline
\raisebox{-18mm}{\includegraphics[width=0.15\textwidth, height=25mm]{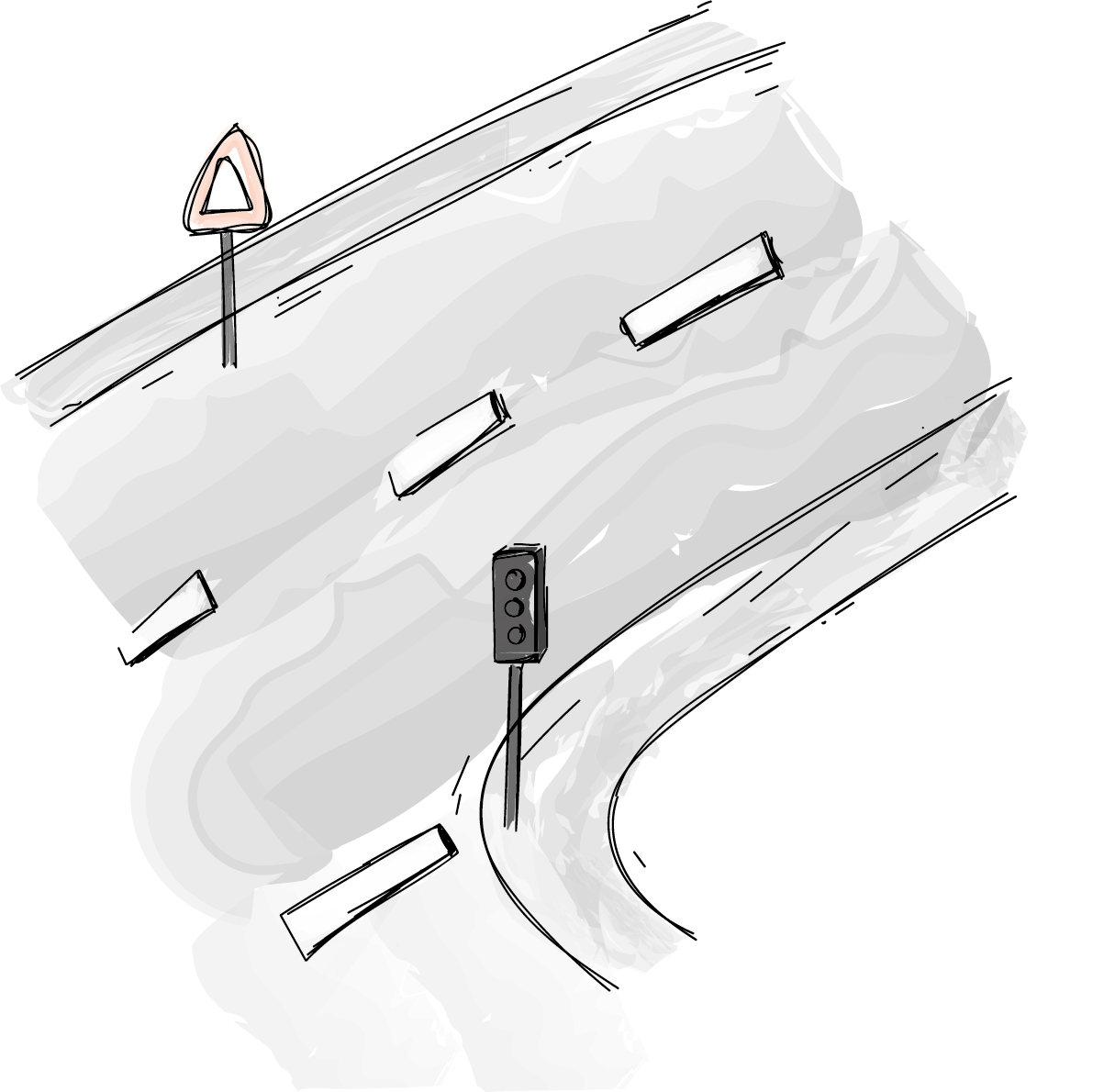}} & \begin{tabular}[c]{@{}l@{}}Roads including shoulders, sidewalks,\\ parking spaces etc.\\ \\ Road markings\\ \\ Traffic signs and traffic lights\end{tabular}                                                                                                   \\ \hline
\end{tabular}
\end{table}

Note that, together with the assignment of the entities to the layers, many properties like position, velocity, size, material or color can, in general, be described within the respective layers. There are, however, also compelling reasons to place certain properties in another layer than the object itself. In some cases, for convenience only, this can be bypassed by using so-called annotations. For more details and guidelines, we refer to Section \ref{sec:Guidelines}.

\subsection{Road Network and Traffic Guidance Objects (Layer~1)}
\label{subsec:L1}

Layer~1 describes the \textit{road network together with all permanent objects required for traffic guidance}. For the definition of `permanent' compare the term `geo-spatially stationary objects' in \cite{b9}, i.e., the objects and properties described in this layer remain unchanged within a scenario. Non-permanent traffic guidance objects are described from Layer~3 upwards. Given the road network and traffic guidance objects, Layer~1 summarizes \textit{where} and \textit{how} traffic participants can drive. 

The road network refers to the geometry, topology and topography of the roads including road course, road linkage, road elevation and lateral profile. With the help of road markings that are located in Layer~1, the semantics of the lanes can be derived, e.g., shoulders, cycle paths, and sidewalks. Furthermore, special areas with their boundaries such as parking spaces and keep-out areas can be clearly marked. In that regard, road markings also comprise instructions that are painted on the road surface like speed limits, stopping lines, or turn arrows which are supplementary to given traffic signs or traffic lights.

Both, permanently present (switchable) traffic signs and traffic lights are placed in Layer~1 while their (changing) states will be described in Layer~6 (see Section \ref{subsec:L6}). Within Layer~1, we assume that the semantics of the traffic signs, traffic lights, and road markings is imposed by the German Road Traffic Act (StVO) and the Vienna Convention on Road Signs and Signals \cite{b25} and thus needs no further explanation. According to the VzKat, the catalog of traffic signs of the StVO \cite{b26}, delineators and beacons like vertical panels with distance information and before railroad crossings belong to the traffic signs and are, therefore, described in Layer~1. However, since many types of beacons, such as vertical panels with integrated warning lights as well as guide barriers and separators, are used exclusively for the delimitation of construction sites, these are no permanent guidance objects and thus not relevant before Layer~3. In contrast to advertising boards and other private signs positioned in subsequent layers, all directional signs - e.g. city limit signs, official tourist signs and direction signs - are placed in Layer~1. With the help of road markings and traffic signs, roundabouts, traffic islands, and bus stops can be identified unambiguously.

Given all this information, it is now easy to plan a traffic participant's mission: start and end point on the road, route to be followed, velocity profile, and handling of, e.g., stopping requests. In addition to the mission planning on this microscopic level, Layer~1 can also be used to design complete traffic flows on a macroscopic level. On that basis, different road networks can be compared and urban ones can be distinguished from rural ones. Moreover, an initial assessment of the complexity of scenarios is already possible on this layer using the information on geometry, topology and topography of the roads as well as traffic guidance objects.
 
Lastly, we assign the road surface material, e.g., asphalt, cobblestone and gravel to Layer~1. It also includes irregularities of the road surface, e.g., damage and potholes. Additional structures such as manhole covers and grids modifying the road surface are described as well. The same holds for speed bumps which are often a combination of street elevation and road markings.

From the board game perspective, Layer~1 is the board on which elements of higher layers can be placed. It extends the definition of the base road network \cite{b11} by including the traffic infrastructure of \cite{b12} and \cite{b13}. To allow for a clear separation from the following layer, however, the focus is on those traffic infrastructure objects which are used for traffic guidance and regulation.

\subsection{Roadside Structures (Layer~2)}
\label{subsec:L2}

In contrast to PEGASUS, current projects like V\&V-Methoden \cite{b27} and SET Level \cite{b28}, as well as research and development in industry, focus on urban environments. Compared to highway scenarios, where the number of objects beyond the road is very limited, there are many roadside objects that need to be described in the urban setting. This justifies a separate layer to subsume all these objects. Our definition of Layer~2 contains many urban structures already considered in the situation-specific adaptations of \cite{b11}.

Layer~2 addresses the \textit{roadside structures} and contains all \textit{static objects} that are usually \textit{placed alongside - and not onto - the road}. Examples of such static objects are buildings, vegetation like trees and bushes, walls and fences, street lamps, above ground hydrants, bollards, and other types of fixed poles. Bus shelters with benches and surrounding constructions like tunnels and bridges are also grouped in this layer. The same holds for so-called vehicle restraint systems that prevent vehicles from leaving the road and reduce crash severity. The safety structures include, for instance, guardrails, concrete step barriers, and impact attenuators. Bridge barriers for cyclists and pedestrians shall also be listed here.

Analogously to Layer~1, we suppose that all objects of Layer~2 are \textit{permanently installed} at a designated position. For reasons of simplicity, however, an oscillating motion of leaves, banners or flags can also described in this layer (see also the guidelines in Section \ref{subsec:Guidlines}). In cases where we need a more detailed motion description or there are deviations from this designated position, we refer to Layer~3 or Layer~4, respectively.

Coming back to the board game example, Layer~2 can be thought of being on top of the basic Layer~1 changing the design of the board and increasing the complexity of the scenario at hand. Keeping simulation applications in mind, this allows to perform simplified or detailed simulations depending on the chosen layers.

\subsection{Temporary Modifications of Layer~1 and Layer~2 (Layer~3)}
\label{subsec:L3}

Layer~3 is comprised of \textit{temporary modifications of elements of Layer~1 and Layer~2}. Therefore, Layer~3 does not introduce any new object classes that have not already been defined in the previous layers. This explicitly does not mean that no new objects can be introduced, just that they are of the \textit{same class as Layer~1 and Layer~2 objects}. An explanatory example is a construction site with corresponding traffic signs and road markings, modifying the course of the lane. Those elements are guidance elements of Layer~1, i.e., of an object class already described, but we assume they are newly introduced in Layer~3. As mentioned in Section \ref{subsec:L1}, beacons within construction sites are included in the catalog for traffic signs. The same holds for traffic cones. Therefore, they are of the class `traffic guidance object' (Layer~1), but are instantiated in Layer~3 as they are \textit{non-permanent} in the real world.

Moreover, a fallen tree (compare also \cite{b29}), boulders, contamination of the road through soil or sand as well as collapsed building parts that are lying on the street (e.g., due to an elementary event) are described in Layer~3. The same holds for the state change of a road marking which is covered by the elements mentioned before. Note that all modifications in Layer~3 \textit{are supposed to be constant for the entire duration of the scenario}.

Given that the scenario duration only represents a lower bound for the word `temporary', a specification of the upper bound seems to be more difficult. In \cite{b12} `temporary' is defined as a time frame of one day, although a thorough evaluation on why this time interval was chosen is not given. To the opinion of the authors, such a justification is necessary. The following paragraph provides possible sources that can serve as orientation on the basis of which the time frame can be chosen.

In order to devise a suitable time frame for temporary modifications and to decide if the choice of one day is justified, different regulations were consulted. These include guidelines for roadwork \cite{b30}, a handbook for road markings \cite{b31} and regulations for temporary buildings \cite{b32}. The German guidelines for roadwork \cite{b30} only differentiate between roadwork of longer duration (minimum one day and stationary) and roadwork of short duration (matter of hours shorter than one day). As already stated, this definition is not practicable when thinking of a holistic environment description, as the time span of one day is too short to be an upper threshold for, e.g., construction work. The handbook for road markings \cite{b31} defines a category for temporary markings that exist longer than 180 days. In general, this is more in accordance with our understanding of possible duration of roadworks and shows that the term `temporary' can also include a rather large time frame.

Temporary building regulations \cite{b32} claim that non-permanent structures are defined as being intended to be disassembled and assembled at different locations. A special permit is necessary, if such an object is located in the same position for more than three months. Therefore, we suggest using this threshold as an orientation for roadside structures.

To conclude, the above shows that existing regulations can only give an indication. `Temporary' cannot be defined independent of the context of the modification and can be a rather flexible notion possibly including large time frames. Some prominent public building projects are perfect examples in this regard. Therefore, it is advisable not to determine a fixed threshold, but to perform this classification on a case by case basis, keeping the specific application of the 6LM in mind. 

\begin{figure*}[th]
\centering
\centerline{\includegraphics[width=0.9\textwidth, scale=0.75]{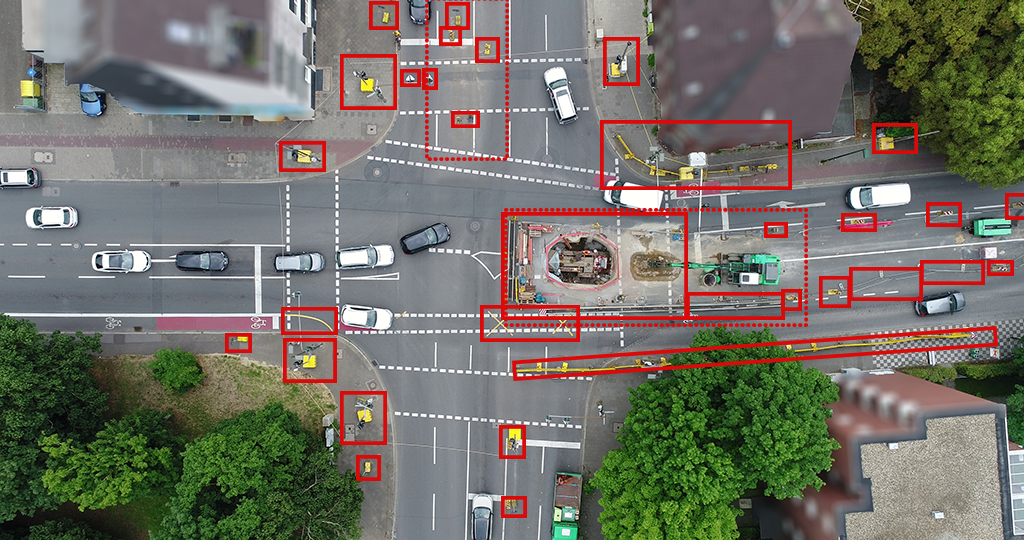}}
\caption{Image recorded using a drone showing an intersection with roadwork. Elements of Layer~3 are marked in red. Boxes drawn by hand for illustration purposes only.}
\label{fig:L3}
\end{figure*}

Fig.~\ref{fig:L3} shows an image of an intersection that is currently under construction. In the figure, entities belonging to Layer~3, i.e., representing temporary changes of Layer~1 and Layer~2 due to the construction site, are highlighted in red. Those are the actual construction site itself, roadwork traffic signs, temporary traffic lights and temporary road markings. It is important to note that the excavator and the little roadwork trailer (on the far right-hand side), which can also be assigned to the construction site, are not part of Layer~3. They are defined as movable, dynamic objects and are, therefore, located in Layer~4.

\subsection{Dynamic Objects (Layer~4)}
\label{subsec:L4}

Layer~4, \textit{`Dynamic Objects'}, is the first layer that introduces a \textit{time-dependent description}. It is roughly speaking the `traffic layer' as it includes \textit{movable objects} whose movements could evolve over time and are described by trajectories or maneuvers. This is consistent with the definition in previous publications within PEGASUS such as \cite{b13} and \cite{b15}. However, when extending the application of the 6LM to the urban context, not only focusing on the SAE Level 3 \cite{b19} functions on highways, many additional entities must be described on this layer.

Layer~4 contains all ``dynamic element'' that, according to \cite{b9}, ``move (having kinetic energy), or possibly being able to move (having sufficient energy and abilities to move)''. The latter refers to the \textit{objects that can potentially move, but do not necessarily have to} within the scenario. These objects might be stationary, resting in a fixed position such as parked vehicles, pedestrians standing still and garbage cans sitting in their pickup-position at the street etc. Note that the definition includes entities that are designed to perform movements, but also comprises those entities that move on a regular basis or due to an external trigger.

The definition in \cite{b9} together with the fact that Layer~4 also describes state changes that are not necessarily associated to the entity's movement led us to a renaming of Layer~4 as `Dynamic Objects' (in contrast to the `Movable Objects' of \cite{b13}). An example for such a `dynamic' (or time-dependent) state change is the visibility of a road marking which is covered by increasing snowfall. In this regard, modifying the original definition of dynamic objects in \cite{b8} and describing them as ``elements whose state changes in-between scenes'' (i.e., within a scenario) complements our understanding of Layer~4.

All traffic participants present good examples for dynamic objects on Layer~4. These include vehicles (including trailers), motorcycles, bicyclists, pedestrians and rail vehicles such as trams. Furthermore, objects on Layer~4 include animals and miscellaneous objects that are movable. 

Let us mention coke cans lying on the road or being kicked by a child and household garbage cans that are generally placed near houses or pushed to and from the street for pick-up. Similar examples are balls rolling towards the street or trees currently falling over. As part of the vegetation, a tree that is planted in a certain position is located on Layer~2. If this tree happens to fall over and find its resting position on the road, this can be understood as a temporary modification and the fallen tree would be situated in Layer~3. However, while the tree is falling over, it features a trajectory and will be placed on Layer~4.

\begin{figure*}[th]
\centering
\centerline{\includegraphics[width=0.9\textwidth, scale=0.75]{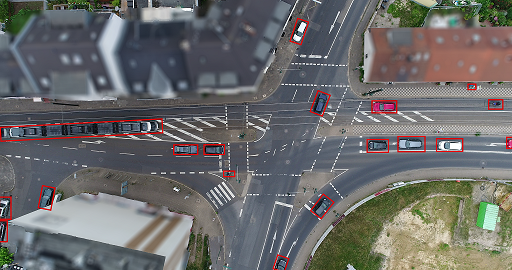}}
\caption{Image of an intersection recorded using a drone highlighting all Layer~4 elements. Boxes drawn by hand for illustration purposes only.}
\label{fig:L4}
\end{figure*}

An image of a regular urban intersection is shown in Fig.~\ref{fig:L4}. In the image, all Layer~4 elements are highlighted through red boxes. It contains vehicles (parked and moving), bicycles and a tram. In the board game example, those are the actors on the board that has been designed through Layer~1 to Layer~3. 

Previous publications \cite{b12} and \cite{b13} included interactions in their description of Layer~4. Considering the overall aim of the 6LM of giving a general environment description, i.e., a description of the physical observable only, the term interactions might be misleading. The 6LM does not include any non-physical entities such as goals and values.
 
In \cite{b8} light and weather conditions are included in the term `dynamic object'. In the case of the 6LM, light and weather conditions are not part of Layer~4 as they are defined in a separate layer, Layer~5, `Environmental Conditions'.

\subsection{Environmental Conditions (Layer~5)}
\label{subsec:L5}
Layer~5 contains \textit{environmental conditions}. These consist of \textit{weather, atmospheric and lighting conditions}. Weather and atmospheric conditions, e.g., include precipitation, visibility (fog etc.), wind, and cloudiness. Layer~5 also includes \textit{road weather conditions}. These are weather related modifications of the road surface like dry, wet or icy roads. As part of the lighting conditions, daytime related aspects such as positioning of the sun (or the moon) and sunrays can be listed. Furthermore, artificial light sources like lighting through street lamps or through advertising boards are included. The current layer also features a \textit{time-based description}, as it is possible that environmental conditions change during a scenario.

By this definition, Layer~5 is consistent with previous publications such as \cite{b13}. However, while the content has not changed, the naming was adjusted to be in accordance with the English term and various corresponding publications describing operational design domains (ODDs) \cite{b34,b35,b36}. 

Note that we define Layer~5 to include globally perceptible environmental conditions. This, e.g., covers whether fog is present or not. Actor-dependent occlusions through, for instance, fog or traffic participants are not covered in the 6LM. They can, however, be derived for different actors from the information given in the 6LM.

Effects of Layer~5 conditions that fall into the category of `globally perceptible conditions' are presented in Layer~3 or Layer~4. If an environmental condition such as wind induces a movement, this must be described in Layer~4 (compare also \cite{b11}). As a consequence, the traffic cone blown away by strong winds or the motion of individual leaves - if desired to describe in detail - are addressed in this layer. Layer~4 also includes a road marking that continuously loses visibility. If the effect of the environmental condition is without any dynamic component and constant for the entire duration of the scenario, we formulate this change in Layer~3. In other words, the road marking which is covered by snow for the entire duration is described in Layer~3. Note that this approach allows the orchestration of information within the 6LM since the categorization of entities and properties into the different layers takes place in an objective way. The observable result is the same even if the triggering event causing it is different: A traffic cone currently being overturned by wind or knocked over by a construction vehicle are both located on Layer~4, a road marking covered by snow or lost cargo is located on either Layer~3 or Layer~4 (depending on whether the road marking's covering state is constant or not).

\subsection{Digital Information Layer~6}
\label{subsec:L6}

Layer~6 is defined to focus on \textit{all kinds of information exchange, communication, and cooperation on basis of digital data only}. A sixth layer was first introduced in \cite{b13} as \textit{`Digital Information'} and later renamed as `Data and Communication' in \cite{b15}. In the publication at hand, we choose the original name.
 
Analogous to previous publications, Layer~6 addresses all information between vehicles, infrastructure, or both, emerging from V2X modules and for instance transmitted wirelessly. V2X is a rather new concept without high market penetration today whose importance will increase with the future developments \cite{b37}. Let us mention information about road closures due to accidents, extreme weather conditions and truck-to-truck communication in platooning applications as examples. Due to its importance for digital information exchange, wireless signal coverage and strength is also assigned to this layer.

In the 6LM, we differentiate between the classification of entities and properties on the one hand and the classification of the underlying information source at the other hand: When describing that a vehicle reduces its velocity due to an upcoming traffic jam, the vehicle's braking action and the traffic jam itself are assigned to Layer~4. Thereby, it does not matter whether the traffic jam was observed by the driver (or a set of sensors in case of automation) or received through a V2X message. If, however, the information was transmitted through V2X, the fact that the V2X message was sent, together with its content, must be represented in Layer~6. We like to stress that this is particularly important when the V2X information differs from the ground truth. Since the reception of V2X messages requires specialized hardware, a placement of V2X messages in Layer~6 is justified as well.

\begin{figure}[th]
\centerline{\includegraphics[scale=0.5]{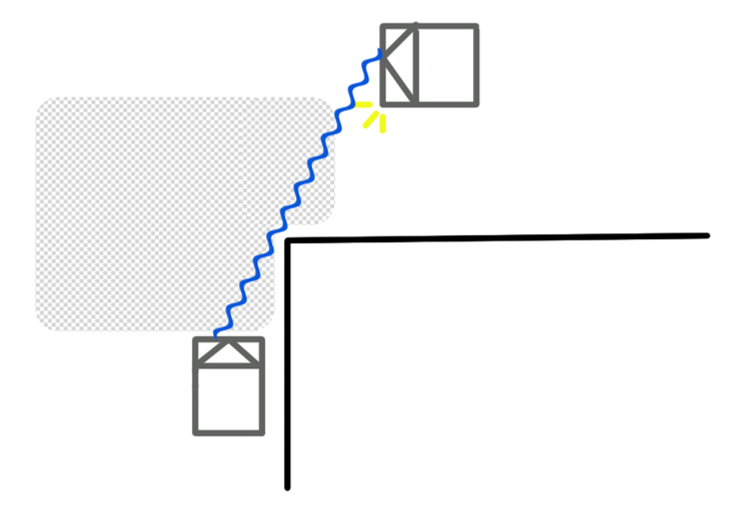}}
\caption{Example for relevant V2V communication at an intersection.}
\label{fig:L6}
\end{figure}

Fig.~\ref{fig:L6} gives an example where the V2V message is supplementary, but identical, to the ground truth information. Two vehicles are approaching an intersection with an occlusion present. The vehicle that has the right of way intends to turn left and, therefore, crosses the planned path of the second vehicle. The intention of turning left can be seen through the indicator light (visible to the other vehicle only without occlusion), but it could also be conveyed through a V2V message. Without the V2V message the environment description would be incomplete. However, for the approaching vehicle that is supposed to yield, the V2V message is relevant due to the occlusion.

Intelligent traffic management systems are assigned to Layer~6, too. Part of traffic management systems are the states of traffic lights and switchable traffic signs as they are, for instance, used for highway sign gantries. Those states are included in Layer~6, no matter if they are controlled centrally for dynamic traffic routing, use V2X communication or feature none of the above. In any case, these states are encoded as digital information. With increasing V2X communication another example are traffic lights that inform the traffic participants about the duration of a green phase and an appropriate velocity to cross the intersection without the necessity to decelerate or stop. Similarly, we can mention the flashing lights at railway crossings. Note that only the variable information, i.e., the changing states of traffic signs and traffic lights, is described here, while the respective objects are already placed in Layer~1.

\section{Guidelines for the 6-Layer Model}
\label{sec:Guidelines}

\begin{figure*}[th]
\centering
\centerline{\includegraphics[width=0.9\textwidth, scale=0.75]{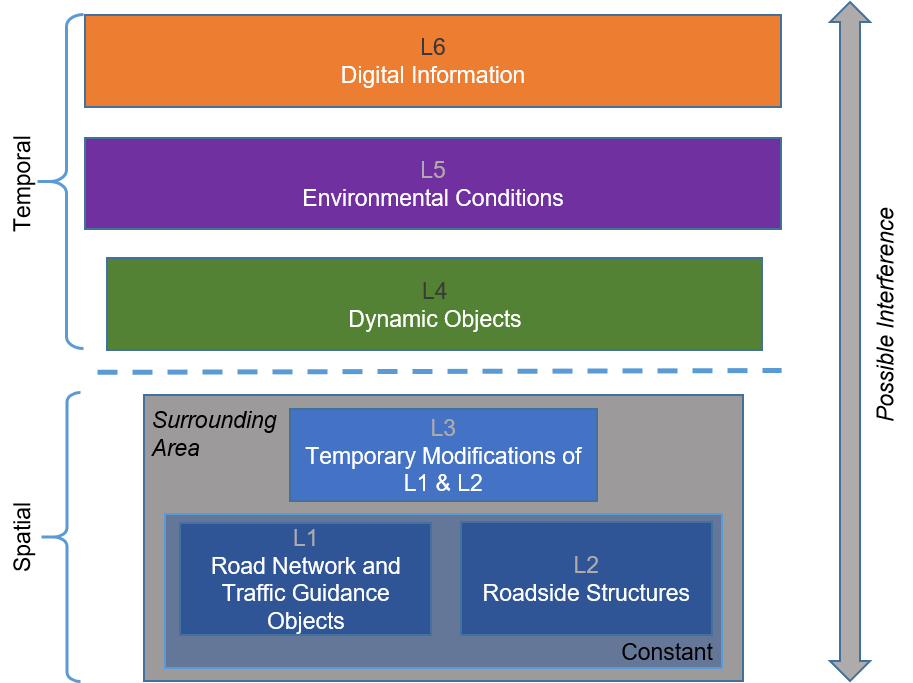}}
\caption{Overview of the layers including spatial and temporal separation.}
\label{fig:Overview}
\end{figure*}

The previous section provided definitions of the layers, their naming and a description of what should be included in the layers. Working with the model as a tool to generate a holistic description of the environment has shown that ambiguities in the process of assigning entities to layers, even though classifications are carefully made. The following guidelines intend to give clarification on this matter and are meant to show that the model is capable of providing an overall categorization. All guidelines are stated first and are subsequently explained along with examples in the following section. Furthermore, Fig.~\ref{fig:Overview} reveals some aspects formulated in the guidelines in a graphical manner.

\subsection{Guidelines}
\label{subsec:Guidlines}

\begin{enumerate}
	\item Layers~1,~2 and 3 conduct a spatial-based description. They do not contain any time-variable aspects. Time-based descriptions are introduced from Layer~4 upwards.
	\item Layer~3 contains temporary changes of Layer~1 and~2. These changes are fixed for the whole duration of the scenario. They are not permanent in the sense of Layer~1 and 2.
	\item From Layer~3 upwards, state changes are introduced. Additionally, from Layer~4 upwards state changes can be time-dependent.
	\item If an entity has time-dependent properties (potentially variable during a scenario), it should be placed on Layer~4 upwards. However, not all its properties need to be time-dependent.
	\item Not all properties of an entity are necessarily in the same layer. The same property of a given entity should, however, not be located on different layers. If in doubt where to locate a property, it is placed in the layer where it matches the description of the layer best and has the largest influence.
	\item Annotations can be used for reasons of simplicity or in order to add extra information.
	\item Allegedly global properties need to be thoroughly checked whether they are truly objective. If they are not, they are not part of the 6LM.
	\item Properties of all layers can influence properties on other layers. There is no single direction of influence.
\end{enumerate}

\subsection{Explanatory Description and Examples for Each Guideline}
\label{subsec:GuidelineExamples}

\textbf{Guideline 1.} The separation of spatial-based and time-based description allows the reusability of Layer~1 and~2 entities if one location is used multiple data recordings. The road network, traffic guidance objects and roadside structures can be kept constant and are only changed by modifications in Layer~3. This separation of spatial description (Layers~1 to~3) and temporal description within a scenario (Layer~4 upwards) is consistent with the separation of entities in the OpenDRIVE \cite{b20} and OpenSCENARIO \cite{b21} description formats \cite{b38}.

\textbf{Guideline 2.} As explained in Section \ref{subsec:L3}, Layer~3 contains temporary changes prevailing for the whole duration of the scenario. As such, objects of Layer~1 and Layer~2 that are modified can be added as well as new objects of classes already contained in Layer~1 and Layer~2.

\textbf{Guideline 3.} In physics, a state describes the collection of all information needed to describe a system at a certain point in time. Then, a state change describes the change of one system state to another system state. If one considers entities within an environment, states can continuously change or remain constant after a single change. The changed state of an object from Layer~1 or Layer~2 is described in Layer~3 if the following holds: A modification to the original state is visible, but the change process itself could not be observed, and the modified state lasts for the duration of the scenario. As an example consider road markings covered by soil for the entire scenario. In the 6LM, only those states whose changes are explicitly observable over time are understood as being time-dependent and thus, described in Layer~4 upwards according to Guideline 2. There can be a single state change or continuous state changes as the road marking can be covered all at once or step by step. Of course, there are also state changes described in Layers~5 and~6, for instance, the change of a weather condition or the state change of a traffic light.

\textbf{Guideline 4.} Guideline 4 requires to locate entities featuring time-dependent changes on Layer~4 upwards. This includes movable objects, as they can change their position. At the same time, however, these entities can also have constant properties, such as size and color, which can be further detailed in an ontology on the basis of the 6LM (see Section \ref{sec:FutureWork}).

\textbf{Guideline 5.} According to the definition of the 6LM and the previous guidelines, it is obvious that there can be properties of a single entity assigned to different layers. A prominent example are traffic lights. Traffic lights are by definition part of the traffic guidance objects and, therefore, located on Layer~1. In Layer~1, they are spatially situated in a specific position. However, traffic lights can also change their states over time by switching, e.g., from green to yellow. This state change is located on Layer~6. This can easily be understood when thinking of the traffic light as two parts: The stationary element and the controller performing state changes (this separation is also in accordance with previous publications, e.g., \cite{b11}). A similar example are street lamps (positioned on Layer~2) that can be switched on and off (state changes in Layer~5 as part of the lighting conditions).

Always locating a specific property of an entity on the same layer is definitely desirable. As such, movements of traffic participants are always mapped to Layer~4 and weather conditions are exclusively addressed in Layer~5. However, there might be examples where the same property is described in different layers. If this is the case, it is due to and in accordance with the definition of the 6LM. Let us, e.g., consider the visibility of a road marking. As stated in the explanation of Guideline 4, it depends on the observability of the state change whether the property is located in Layer~3 or Layer~4.

We provided detailed and concise definitions for the 6LM. Nevertheless, we acknowledge that ambiguous situations may arise and would also like to give a guideline in case of doubt. Anticipating the function of an entity or property within a scenario and using it for the categorization cannot be done as the 6LM is an unbiased description that does not contain any interpretation. If it is not clear where to place an object or property, we will choose the layer where the object or property has the largest influence. Influences on other layers are, however, still possible. 

The control and state change of a traffic light will be located in Layer~6 as part of the `Digital Information' Layer. They are not located in Layer~5 although one could argue that the state change, e.g., from red to green, might influence the lighting condition of the surroundings. That being said, the influence of the state as traffic regulation element is rated much higher and is, therefore, placed in Layer~6.

\textbf{Guideline 6.} Annotations can be used for two reasons if desired. On the one hand, they can be used to add extra - interesting, but not necessarily essential - information regarding entities. Prominent examples for this are annotations to emergency responders. A police officer is described on Layer~4, as he is a pedestrian. However, we might want to annotate that he is on duty fulfilling some regulatory task. Similarly, the fact that emergency vehicles are executing their privileges through siren or blue lights can be annotated with the entity on Layer~4.

On the other hand, annotations can be used for reasons of simplicity when an information is valuable, but does not need to be given in detail. One example for this was already mentioned in the definitions of Layer~2 and Layer~5 when describing the motion of leaves. In general, an object for which we want to describe motion needs to be placed in Layer~4. This is, e.g., obvious for traffic participants. Thinking of a bush, however, it might be desirable to only note that it is moving in the wind, but it is not necessary to describe the movement of each leaf in detail on Layer~4. Therefore, the general information of an oscillating motion can also be annotated with the object on Layer~2. Analogously, warning lights at railroad crossings or on safety beacons can be flashing. If desired, this state change can be described on Layer~6. However, if the state does only undergo simple periodic changes (`flashing') this can also be marked as an annotation to the entity itself on the same layer the entity is originally placed.

\textbf{Guideline 7.} A prominent example for an allegedly global property is the friction coefficient. On first sight, it would be possible to place the friction coefficient with the road or the environmental conditions. However, when considering properties that influence the friction coefficient, it becomes clear that it does not only depend on the road surface (asphalt, cobblestone, gravel, etc.) and the condition of the road (dry, wet, icy, etc.), but also on the material that is in contact with the road. In case of the vehicle that would be the tire. At this point, it becomes clear that the description is not a global and objective part of the general environment description anymore. Therefore, such values are not part of the 6LM. It is, however, possible to determine such values by using several properties present in the description.

Another striking example for the described actor-independence might be occlusions. The 6LM itself does not describe the occlusion for any traffic participant, but it gives the possibility to perform an objective and complete environment description that allows to calculate such occlusions for individual instances of interest in a later step.

\textbf{Guideline 8.} Each layer can influence previous layers and following ones. This idea can also be found in \cite{b12}. The possibility of influence is independent from the numbering of the layers which are not ranked by importance, but purely for structuring of the categorization. For instance, dense traffic could cause that the shoulder is used for driving. This would be an influence of Layer~4 onto Layer~1 that is recorded in Layer~3. Other examples are the traffic light states that influence the trajectories / maneuvers of traffic participants and the weather conditions having an influence on the visibility of road markings. 

With the guidelines presented in this section, the authors hope to facilitate the use of the 6LM and to encourage users from research and industry to apply the model as a tool for structured environment description.

\section{Evaluation through Real-World Data}
\label{sec:Evaluation}
The previous sections gave a definition of the different layers of the 6LM and provided guidelines for a clear classification. This section applies the definitions and guidelines to a real-world example in order to show the practicability of the 6LM.

\begin{figure*}[th!]
\centering
\centerline{\includegraphics[width=\textwidth, scale=0.75]{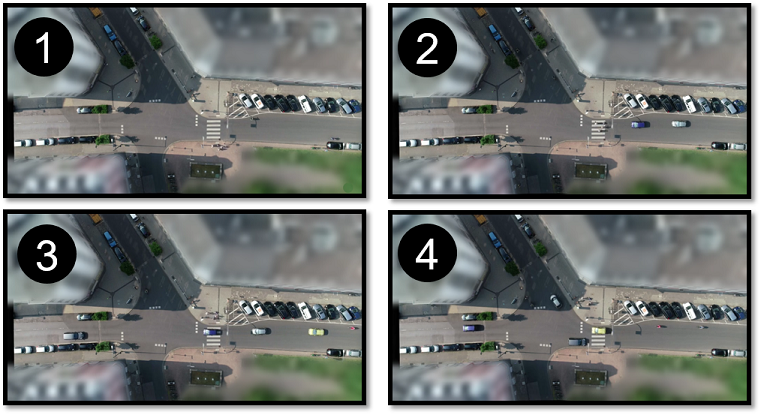}}
\caption{Time sequence of snapshots from real-world data recorded at an intersection by utilizing a drone.}
\label{fig:Scenario}
\end{figure*}

An environment description is given for a real-world measurement recorded by a drone taken from the Intersection Drone (inD) - Dataset \cite{b33}. Fig.~\ref{fig:Scenario} shows four frames out of the video footage for which we analyze the physical occurrences. While each of the frames establishes a scene, a sequence of them is printed here showing the temporal evolvement, i.e. the scenario. The recording features an urban, four-armed intersection. Buildings are blurred in the recordings in order to meet regulations on privacy of information.

In the following, the visible elements in the scenes presented in Fig.~\ref{fig:Scenario} are listed according to the 6LM. Please note that all described elements of Layer~1 to Layer~4 feature descriptions of their material. Those go hand in hand with the object itself and are put into the same layer.

\noindent\textbf{Layer~1}
\begin{itemize}
	\item Geometry and course of the road (including sidewalks)
	\item Traffic signs (hardly visible in Fig.~\ref{fig:Scenario})
	\item All markings: stop lines, crosswalk, markings for parking zones and keep-out areas
	\item Road surface with irregularities
\end{itemize}

\noindent\textbf{Layer~2}
\begin{itemize}
	\item Street lamps (e.g. next to the crosswalk)
	\item Buildings and roadside structures, such as fountains (bottom middle), bicycle stands and vegetation
\end{itemize}

The presented recording does not contain any content for \textbf{Layer~3} as no temporary modifications of Layer~1 and Layer~2 elements are made. After considering Layers~1,~2 and~3 the description of the spatial and non-temporal properties is complete. The description made on Layer~1 and Layer~2 remains invariant when further real-world data of this intersection is recorded. This is consistent with the modeling in OpenDRIVE \cite{b20} and an advantage of the clear separation between spatial and temporal properties that is established in the paper.

From Layer~4 upwards a time-based description is introduced. The temporal development of the events is revealed through the series of images in Fig.~\ref{fig:Scenario}.

\noindent\textbf{Layer~4}
\begin{itemize}
	\item All vehicles, moving and non-moving: The four vehicles on the street as well as the parked vehicles
	\item All bikes: driving on the street and parked at the bicycle stand
	\item All pedestrians, moving and non-moving, e.g., the group of four pedestrians using the crosswalk
\end{itemize}

Without enriching the recorded data with additional information, obtaining a complete description on \textbf{Layer~5} is difficult. Apparent in the recording is that no precipitation or harsh weather condition such as low visibility is present. Therefore, as part of the road weather the street can be described as dry. Furthermore, shadows are visible for different objects. Those shadows are part of the lighting conditions on Layer~5.

This intersection does not feature any traffic lights or switchable traffic signs, therefore, there is no requirement to depict their status on \textbf{Layer~6}. In the same way any other type of Layer~6 information is also difficult to recover from the real-world data recording. Cellular network coverage could be gathered through data enrichment or actual measurements along with the recording of the video footage. Additionally, the recorded intersection is not equipped with any V2X infrastructure and, to the best of our knowledge, there are no traffic participants featuring this technology. Therefore, there is no digital information present

\section{Future Work}
\label{sec:FutureWork}

In order to achieve a standardized description, the authors plan to implement the 6LM as part of a domain ontology for the verification and validation of highly automated vehicles, including a taxonomy of relevant traffic entities as well as their properties and relations. According to the given classification of the 6LM, this ontology will detail the different entities, properties and relations in a comprehensive way, e.g., by introducing subclasses for traffic participants and adding specific weather conditions.

Further, it can be helpful to distinguish between environment description for driving functions and machine perception. Factors triggering actions within a scenario, which might be worth investigating, can be very different for the two groups. Furthermore, when considering perception related aspects the current 6LM might need some adaptations or add-ons as the description of perception aspects cannot necessarily be made actor-independently. Future work will thus concentrate on possible environment description for machine perception. For instance when looking at driving functions, it is often sufficient to look at road surface material. When looking at perception aspects material descriptions for several other entities increase in importance. Moreover, occurrences of reflection and contamination \cite{b39} become of interest. Therefore, it is then necessary to give much more detailed descriptions of all surroundings, not only of the road surface. This could possibly introduce new properties into the 6LM or require a more detailed description of already mentioned properties. This issue could also be addressed in the aforementioned domain ontology.

Additional future work will deal with possible data sources to enrich recorded test data.

\section{Summary}
\label{sec:Summary}

The categorization and guidelines given in this work refine the 6LM for environment description that was originally established in PEGASUS. Furthermore, the model is extended to serve a variety of new applications from verification and validation and to address more complex scenarios than just highway scenarios. Applying the 6LM to urban environments required integrating a concept for roadside structures, other types of dynamic objects and traffic light states.

The definitions and guidelines given help to gain a standardized categorization for a generally usable, unbiased and objective environment description. The work covers all layers and gives explanatory examples along with guidelines. The clear structure established in the 6LM can be used as a basis for scenario descriptions and ontologies.

\section*{Acknowledgment}

The research leading to these results is funded by the German Federal Ministry for Economic Affairs and Energy within the project `VVM - Verification \& Validation Methods for Automated Vehicles Level 4 and 5'.

\begin{IEEEbiography}[{\includegraphics[width=1in,height=1.25in,clip,keepaspectratio]{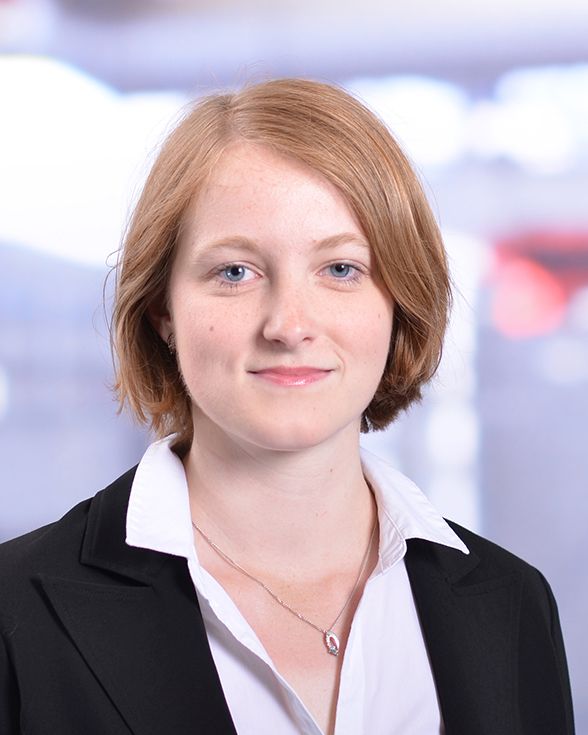}}]{Maike Scholtes} received her B.Sc. and M.Sc. (Dean's List Award) degrees in Computational Engineering Science from RWTH Aachen University, Germany in 2015 and 2017, respectively. During semesters abroad she, inter alia, studied at the University of California, San Diego. After working one year in the area of driving simulation at fka SV Inc. in the Silicon Valley, she started her Ph.D. at the Institute for Automotive Engineering (ika) at RWTH Aachen University. She works on the assessment of automated driving and ADAS with a focus on machine perception.
\end{IEEEbiography}

\begin{IEEEbiography}[{\includegraphics[width=1in,height=1.25in,clip,keepaspectratio]{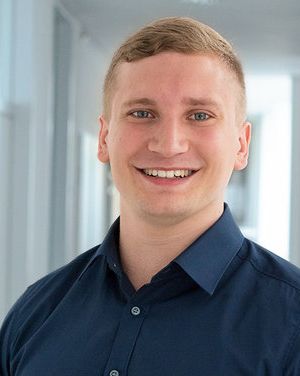}}]{Lukas Westhofen} received the degrees of B.Sc. and M.Sc. (with honors) in computer science by the RWTH Aachen University in 2015 and 2019, specializing on the topics of probabilistic programs and software verification. Since 2019, he is a PhD student at OFFIS e.V., Oldenburg, Germany. His general work focuses on developing methods to establish confidence in the safety of automated vehicles. More specifically, his research interests include the formalization of knowledge as well as its exploitation for safeguarding automated driving functions.
\end{IEEEbiography}

\begin{IEEEbiography}[{\includegraphics[width=1in,height=1.25in,clip,keepaspectratio]{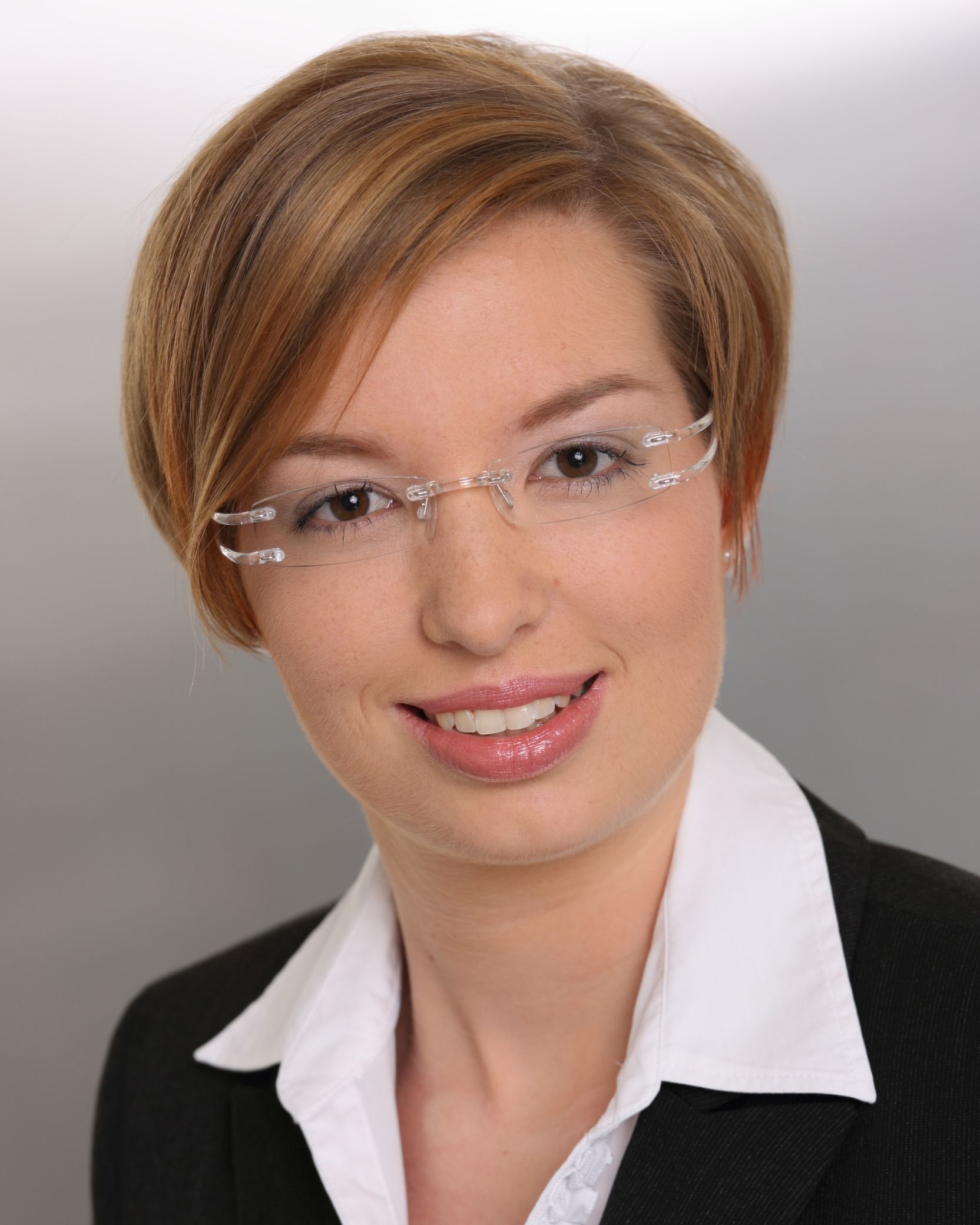}}]{Lara Ruth Turner} received both her diploma degree and Ph.D. in mathematics from the University of Kaiserslautern in 2008 and 2012, respectively. After a postdoctorate at the University of Vienna, she joined ZF Friedrichshafen AG in 2014. She currently works on validation for assisted and autonomous driving functions with a special focus on scenario-based methods.
\end{IEEEbiography}

\begin{IEEEbiography}[{\includegraphics[width=1in,height=1.25in,clip,keepaspectratio]{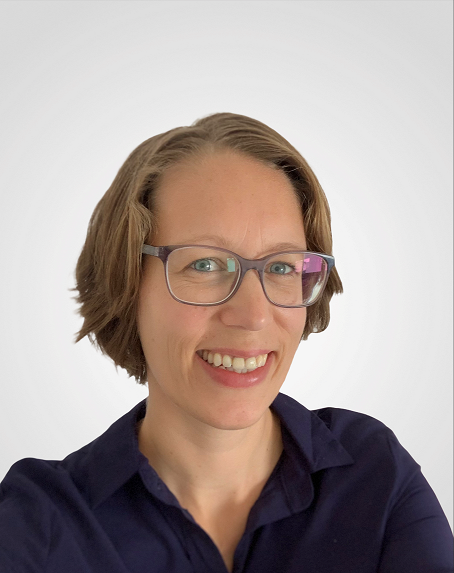}}]{Katrin Lotto} graduated with diploma in mathematics at Technische Universität München, Germany. Since 2009 she has been working at ZF Group in Friedrichshafen, Germany, in various areas of transmission development to operate now in the field of validation methods for automated vehicles. 
\end{IEEEbiography}

\begin{IEEEbiography}[{\includegraphics[width=1in,height=1.25in,clip,keepaspectratio]{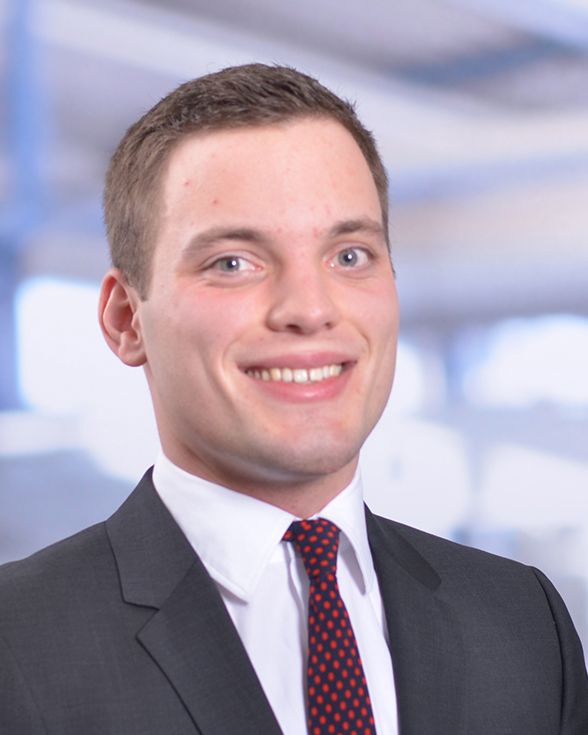}}]{Michael Schuldes} received the B.S. and M.S. degree in electrical engineering from RWTH Aachen University, Germany, in 2016 and 2019 respectively. He is currently a Research Assistant with the Institute for Automotive Engineering, RWTH Aachen. His research interest lies in the area of assessment of autonomous driving functions with a focus on data-driven scenario-based approaches.
\end{IEEEbiography}

\begin{IEEEbiography}[{\includegraphics[width=1in,height=1.25in,clip,keepaspectratio]{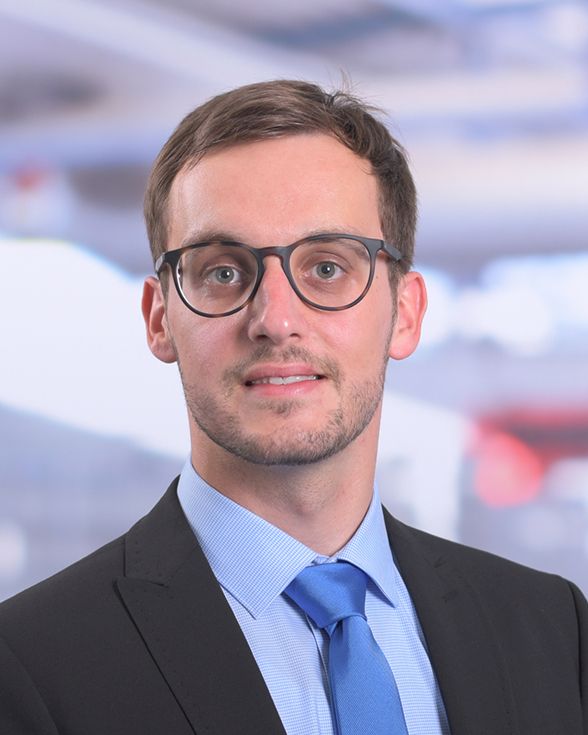}}]{Hendrik Weber} received his B.Sc. in Mechanical Engineering in 2015 and his M.Sc. in Automation Engineering in 2017, both from RWTH Aachen.
Since 2017, he has been a research assistant at the Institute for Automotive Engineering (ika) at RWTH Aachen University, where he is also pursuing his PhD.
His work focusses on verification and validation approaches for automated vehicles as well as on methods and tools for assessing their impact on safe traffic. Currently, he is leading the evaluation subproject in the European research project L3Pilot, which pilots L3 vehicles on public roads and evaluates their behavior in traffic and resulting impacts.
\end{IEEEbiography}

\begin{IEEEbiography}[{\includegraphics[width=1in,height=1.25in,clip,keepaspectratio]{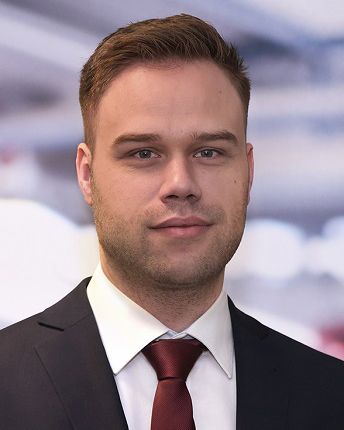}}]{Nicolas Wagener} received the B.Sc. and M.Sc. degrees in Computer Science from RWTH Aachen University, Germany, in 2014 and 2016, respectively. Since the end of 2016, he is a research scientist and PhD student at the Institute of Automotive Engineering (ika) at RWTH Aachen University. Currently, he is Group Leader for Simulation within the research area Vehicle Intelligence \& Automated Driving at the Institute. Moreover, he is the scientific coordinator of the sub-project ‘Databases’ within the project VVMethods. His research interest include simulations and driving simulators, safety validation methods and automated driving.
\end{IEEEbiography}

\begin{IEEEbiography}[{\includegraphics[width=1in,height=1.25in,clip,keepaspectratio]{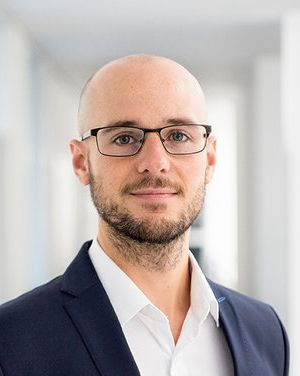}}]{Christian Neurohr} received the B.Sc. and M.Sc. degrees in mathematics in 2011 and 2013 respectively, both from Technische Universit\"at Kaiserslautern, Germany and the Ph.D. degree (Dr. rer. nat.) from Carl von Ossietzky Universit\"at Oldenburg, Germany in 2018. After a short period as a visiting researcher with the MAGMA group at the University of Sydney, he started his current occupation as a postdoctoral researcher at OFFIS e.V., Oldenburg, Germany, where he is working in the area of scenario-based verification and validation of automated vehicles. Specifically, he is the scientific coordinator of the sub-project 'Criticality Analysis' within the project VVMethods.
\end{IEEEbiography}

\begin{IEEEbiography}[{\includegraphics[width=1in,height=1.25in,clip,keepaspectratio]{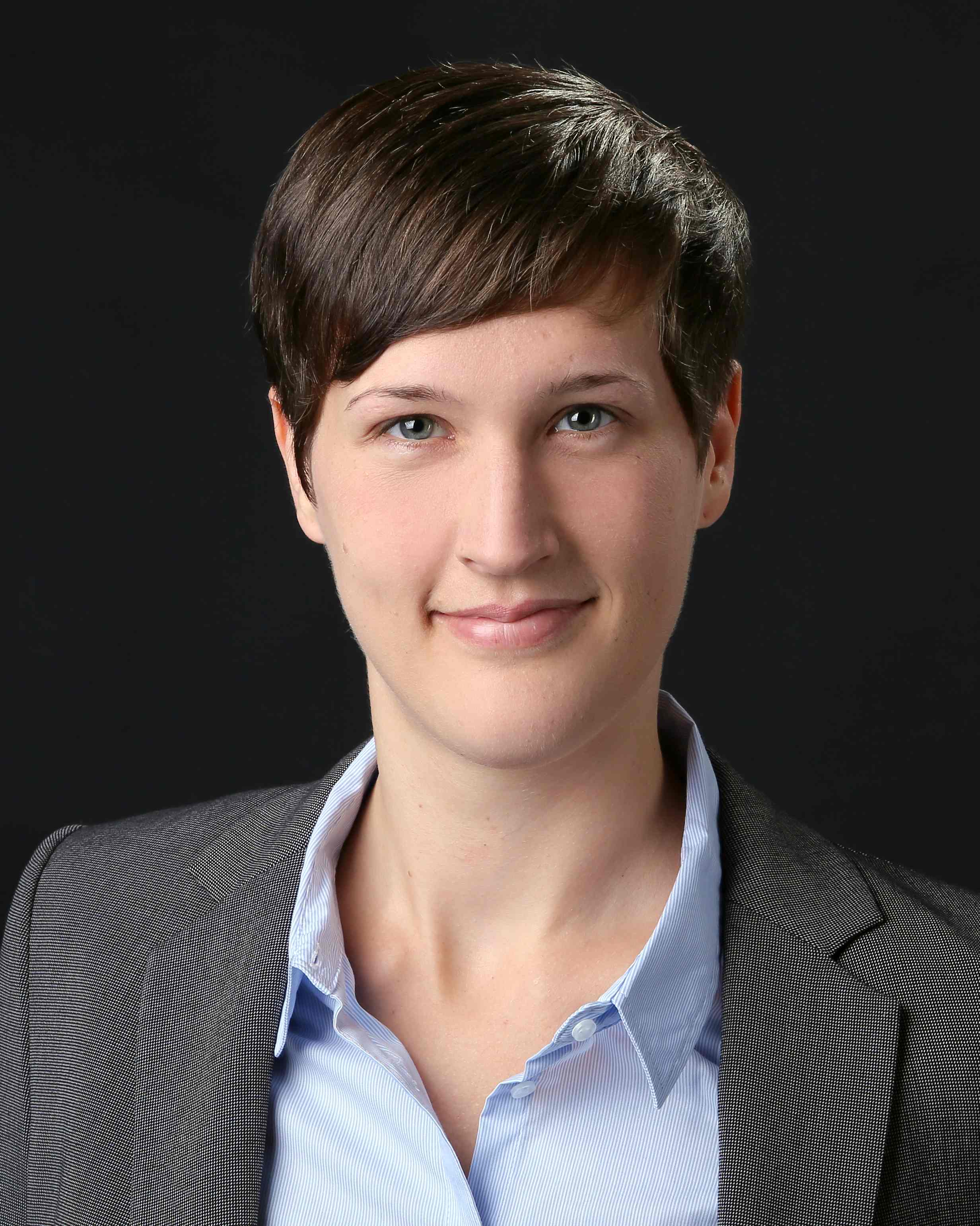}}]{Franziska K\"ortke} graduated with diploma in mechanical engineering at the Technical University of Dresden, Germany. Since 2017, she has been working in the field of test automation for scenario-based validation of automated vehicles at ZF Group in Friedrichshafen, Germany.
\end{IEEEbiography}

\begin{IEEEbiography}[{\includegraphics[width=1in,height=1.25in,clip,keepaspectratio]{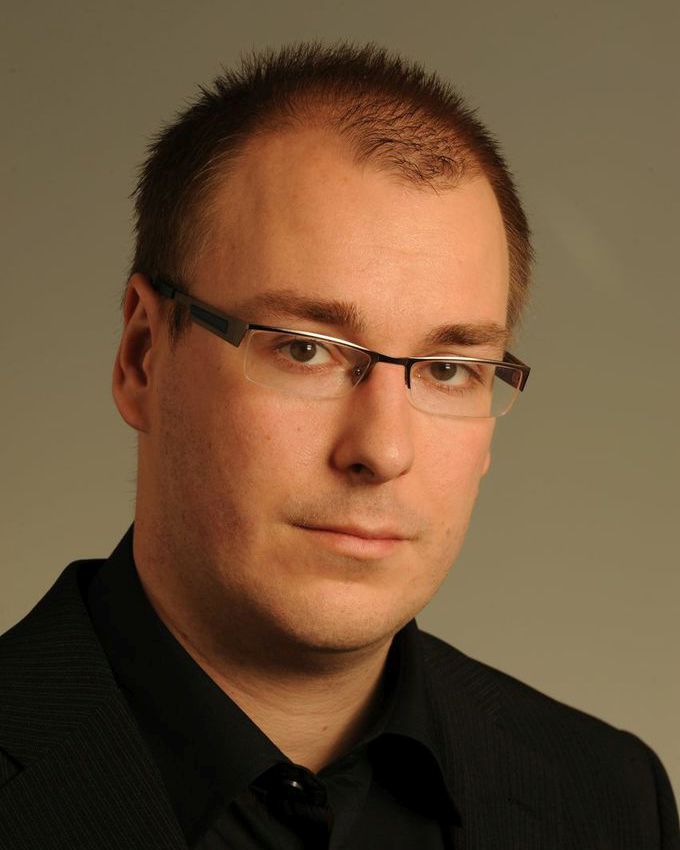}}]{Martin Herbert Bollmann} received the Dipl.-Ing. in mechanical engineering from Technical University Dresden in 2012. Until 2019, he worked as a calculation and reliability engineer at ZF Group in Friedrichshafen, Germany. Currently, he works as test architect, developing test strategies and searching for test methods, that are aligned with SOTIF and able to fulfill requirements coming from safety-related standards. Moreover,
he is the industrial coordinator of the sub-project 'Criticality Analysis' within the project VVMethods.
\end{IEEEbiography}

\begin{IEEEbiography}[{\includegraphics[width=1in,height=1.25in,clip,keepaspectratio]{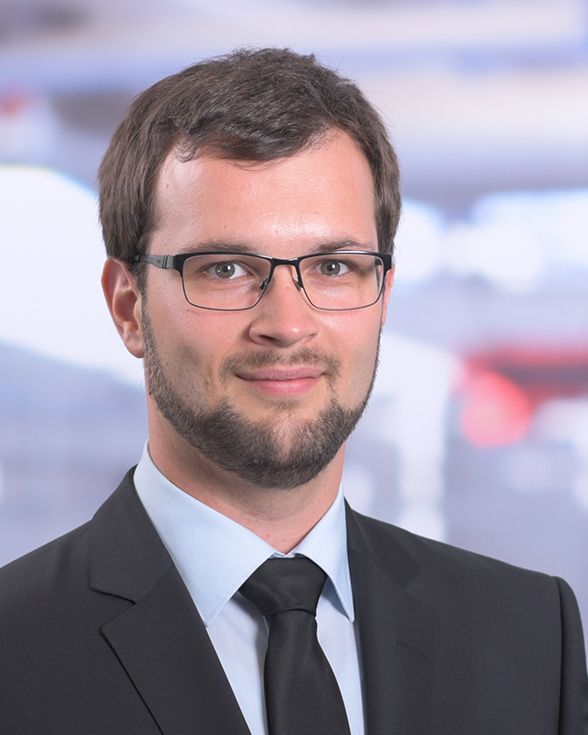}}]{Johannes Hiller} studied Electrical Engineering, Information Technology and Computer Engineering at RWTH Aachen University. Currently, he is Group Leader Data \& Intelligent Infrastructure within the research area Vehicle Intelligence \& Automated Driving at the Institute for Automotive Engineering (ika) at RWTH Aachen University. He works on the assessment and evaluation of automated driving and advanced driver assistance systems with a focus on data analysis, enrichment and the analysis of video.
\end{IEEEbiography}

\begin{IEEEbiography}[{\includegraphics[width=1in,height=1.25in,clip,keepaspectratio]{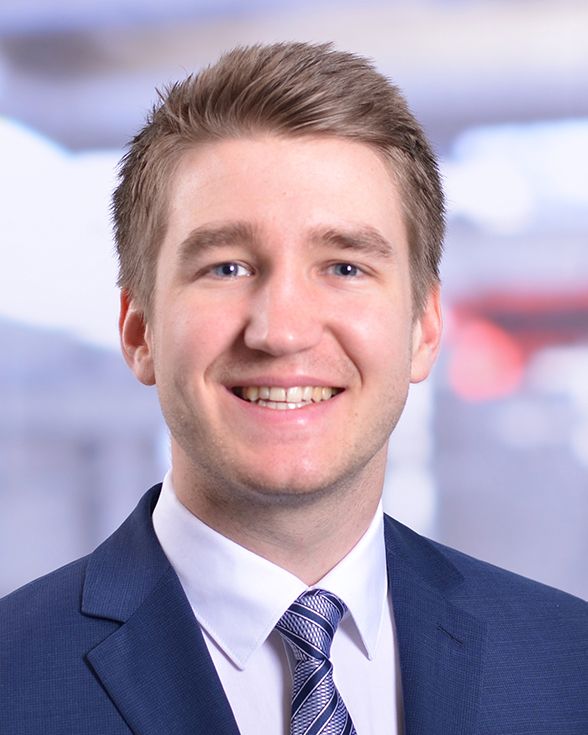}}]{Michael Hoss} received the B.Sc. and M.Sc. degrees in Computational Engineering Science from RWTH Aachen University, Germany, in 2015 and 2017, respectively. Since the beginning of 2018, he has been a research associate at the Institute for Automotive Engineering (ika) of the same university, where he is also pursuing a Ph.D. degree. His research interest focuses on safety-aware test methods for the perception subsystem. 
\end{IEEEbiography}

\begin{IEEEbiography}[{\includegraphics[width=1in,height=1.25in,clip,keepaspectratio]{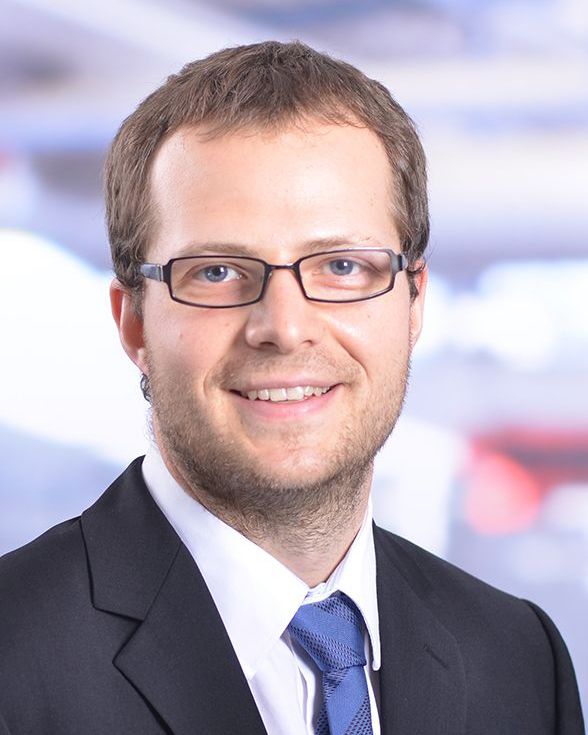}}]{Julian Bock} received the B.Sc. degree in Computational Engineering Science from RWTH Aachen University in 2012 and the M.Sc. degree in Computational Engineering Science in 2014. From 2014 to 2019, he was research scientist and PhD student at the Institute for Automotive Engineering at RWTH Aachen University.  Since 2020, he is Manager Artificial Intelligence at fka GmbH in Aachen. His research interests include automated driving and its fields safety validation, real-world scenario data acquisition, machine learning and pedestrian prediction.
\end{IEEEbiography}

\begin{IEEEbiography}[{\includegraphics[width=1in,height=1.25in,clip,keepaspectratio]{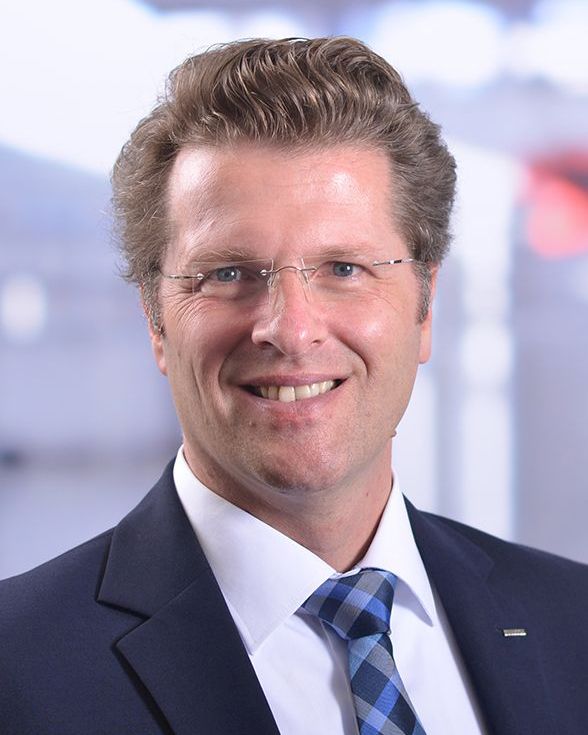}}]{Lutz Eckstein} is head of the Institute for Automotive Engineering (ika) at RWTH Aachen University. He studied mechanical engineering at Stuttgart University and received his Ph.D. from the same university in the year 2000. During his time in the industry he worked on active safety systems and human-machine interactions for Daimler AG and BMW AG. He was appointed to his professorship and current position at RWTH Aachen University in 2010.
\end{IEEEbiography}

\EOD

\end{document}